\documentclass[acmsmall, nonacm]{acmart}

\usepackage{booktabs,tabularx,longtable, array}
\usepackage{makecell} 
\usepackage{caption}
\usepackage{subcaption}
\usepackage[vskip=1em,font=itshape,leftmargin=2em,rightmargin=2em]{quoting}
\usepackage{xspace}
\usepackage{multirow}

\newcommand{\pquote}[1]{{\textit{\textcolor{black}{``#1''}}}}

\newcommand{\dquote}[1]{{``#1''}}

\newcommand{\added}[1]{\textcolor{black}{#1}}
\newcommand{\deleted}[1]{\textcolor{lightgray}{}}

\usepackage{listings}
\usepackage{xspace}
\usepackage{tikz}
\usetikzlibrary{calc}

\usepackage{marginnote}

\AtBeginDocument{%
  \providecommand\BibTeX{{%
    \normalfont B\kern-0.5em{\scshape i\kern-0.25em b}\kern-0.8em\TeX}}}

\renewcommand\footnotetextcopyrightpermission[1]{} %
\begin{document}


\title{How can LLMs Support Policy Researchers? Evaluating an LLM-Assisted Workflow for Large-Scale Unstructured Data.}


\author{Yuhan Liu}
\affiliation{%
  \institution{Princeton University}
  \city{Princeton}
  \state{New Jersey}
  \country{USA}
  \postcode{08540}
}
\email{yuhanl@princeton.edu}

\author{Shuyao Zhou}
\affiliation{%
  \institution{Princeton University}
  \city{Princeton}
  \state{New Jersey}
  \country{USA}
  \postcode{08540}
}
\email{sz8740@princeton.edu}

\author{Jakob Kaiser}
\affiliation{
\institution{Nuremberg Institute for Market Decisions}
\city{Nuremberg}
\country{Germany}
}

\author{Ella Colby}
\authornotemark[1]
\affiliation{%
  \institution{Princeton University}
  \city{Princeton}
  \state{New Jersey}
  \country{USA}
  \postcode{08540}
}
\email{ellacolby@princeton.edu}

\author{Jennifer Okwara}
\authornotemark[1]
\affiliation{%
  \institution{Princeton University}
  \city{Princeton}
  \state{New Jersey}
  \country{USA}
  \postcode{08540}
}
\email{jo4986@princeton.edu}

\author{Maggie Wang}
\authornote{Authors contributed equally to this research.}
\affiliation{%
  \institution{Princeton University}
  \city{Princeton}
  \state{New Jersey}
  \country{USA}
  \postcode{08540}
}
\email{maggiewang@princeton.edu}

\author{Varun Nagaraj Rao}
\affiliation{%
  \institution{Princeton University}
  \city{Princeton}
  \state{New Jersey}
  \country{USA}
  \postcode{08540}
}
\email{varunrao@princeton.edu}

\author{Andr{\'e}s Monroy-Hern{\'a}ndez}
\affiliation{%
  \institution{Princeton University}
  \city{Princeton}
  \state{New Jersey}
  \country{USA}
  \postcode{08540}
}
\email{andresmh@princeton.edu}

\begin{abstract}
Policy researchers need scalable ways to surface public views, yet they often rely on interviews, listening sessions, and surveys—analyzed thematically—that are slow, expensive, and limited in scale and diversity. LLMs offer new possibilities for thematic analysis of unstructured text, yet we know little about how LLM-assisted workflows perform for policy research. Building on a workflow for LLM-assisted thematic analysis of online forums, we conduct a study with 11 policy researchers, who use an early prototype and see it as a quick, rough-and-ready input to their research. We then extend and scale the workflow to analyze millions of Reddit posts and 1,058 chatbot-led interview transcripts on a policy-relevant topic, treating these sources as rich and scalable data for policy discourse. We compare the synthesized themes to those from authoritative policy reports, identify points of alignment and divergence, and discuss what this implies for policy researchers adopting LLM-assisted workflows.
\end{abstract}





\maketitle
\renewcommand{\shortauthors}{Liu et al.}
\section{Introduction}

Policymakers need timely and nuanced insight into how people experience and talk about policy issues. They routinely draw on public opinion research, by \textit{policy researchers}, to inform decisions, design programs, and evaluate trade-offs~\cite{moon1999opinion, hillygus2011evolution, brewer2024parties, nancarrow2004polls, rasmussen2018little, burstein2014american, stimson2018public}. Policy researchers typically work with three main kinds of data sources~\cite{engler2020all}: \emph{primary} sources such as surveys, interviews, listening sessions, and focus groups; \emph{secondary} sources such as government statistics (e.g., Bureau of Labor Statistics, National Institutes of Health in the U.S.) and think tank reports (e.g., Pew Research); and \emph{microsimulations} that model likely impacts of policy changes. These methods are well-established and crucial when policy researchers need population-level estimates and demographic breakdowns~\cite{britannica_opinion_poll, berinsky2017measuring, simmons1993capturing}. Yet they are slow and expensive to run, require substantial coordination between participants and human researchers, and often reach only people who have the time and resources to participate~\cite{landemore2021opendemocracy, jacobs2011oxford, schulman2024methods}. As a result, policy memos and briefs may miss perspectives from underserved communities, even as initiatives such as the U.S. Office of Management and Budget’s Public Participation and Community Engagement effort call for more diverse voices in policy processes~\cite{schulman2024methods}.

In parallel, online communities have become a major venue where people discuss policy-relevant issues in their own words. Platforms like Reddit host candid conversations about technologies, work, health, and public services, often including detailed anecdotes and critical viewpoints that can be hard to surface in structured surveys or formal consultations~\cite{nagaraj2025rideshare, fiesler2024remember, li2021developers, kim2025capturing, dogan2024narrative, nagaraj2025rideshare, knittel2021anyone}. These discussions are abundant, relatively inexpensive to collect, and already organized into topical communities, making them an attractive potential input for policy research. However, at the same time, social media data raise well-known concerns around representativeness, mis- and dis-information, consent, and platform governance. User populations are skewed, some groups are underrepresented, and access constraints and ethical considerations limit what can be collected and reused~\cite{ferrer2021discovering}. As a result, it is unclear when and how analyses of online discussions can complement more conventional methods, such as targeted interviews, in ways that are useful and responsible for policy work.

Recent progress in LLMs offers new tools for working with these growing text corpora. LLMs can help researchers organize and summarize large-scale unstructured text, including extracting topics and themes from interviews, forums, and other qualitative sources~\cite{pham2024topicgpt, lam2024concept, shankar2024docetlagenticqueryrewriting, rao2024quallm, nagaraj2025rideshare}. Rather than replacing qualitative methods, LLM-based workflows can reduce manual coding effort and make it feasible to incorporate more and larger datasets into early-stage sensemaking. For example, QuaLLM is one such LLM-assisted framework that uses a multi-stage prompting pipeline to extract themes from unstructured text~\cite{rao2024quallm}. These prior works show that this kind of workflow can structure online discussions while keeping researchers in the loop. But we know little about how it fits into policy research practice, how policy researchers experience it, or how its outputs compare to the human-authored reports that currently guide policymakers. Moreover, existing systems are often code-centric and were not designed around the needs of non-programmer policy researchers.

In this paper, we adapt and extend a QuaLLM-style workflow for the specific needs of policy researchers and evaluate it in two studies. Our central research question is: \emph{how can an LLM-assisted thematic analysis workflow help policy researchers make sense of large-scale unstructured text, and how do its outputs relate to the authoritative reports they already trust?} We first focus on policy researchers as users of such a tool. In \textbf{Study~1}, we implement a user interface on top of the workflow that allows researchers: (1) select relevant data sources (currently Reddit communities), (2) define high-level themes, and (3) visualize structured reports with subtopics and example quotes. We ask 11 experienced policy researchers to use the interface to explore two policy-relevant topics (social media use by minors and climate change), compare it to their own non-AI research approach under time constraints, and reflect on where such a workflow fits into their practice. 

 In Study 1, we find that some policy researchers are skeptical about using social media data, in part because platforms are not representative of the wider population and because forum discussions tend to cluster around specific communities and concerns. To examine these issues, \textbf{Study~2} holds the policy topic constant—the economic impact of AI—and compares thematic analysis results across two sources. On one track, we scale the workflow to identify relevant data from Reddit in a large and comprehensive scale. Specifically, we filter from 25,691 subreddits, extract 122,191 quotes from 5,491,991 associated posts. We then apply the workflow to generate themes about AI’s economic impacts. On the other track, we conduct 1,058 semi-structured, chatbot-led interviews with a demographically diverse sample of U.S.\ adults and apply the same workflow to the resulting transcripts. We then compare the themes from both sources to the themes extracted from authoritative policy reports on AI and the economy.

Across both studies, we position the LLM-assisted workflow as a complement to, not a replacement for, traditional policy research methods, particularly for early-stage topic exploration. Our findings suggest that the workflow can recover many themes emphasized in authoritative reports, surface additional community-specific and early-emerging concerns, and help policy experts move efficiently from unstructured text to structured overviews—while also revealing important gaps related to metadata, representativeness, and trust in AI-generated summaries. In summary, the work makes three contributions:

\added{\begin{itemize}
    \item We demonstrate how an existing LLM-assisted thematic analysis workflow can be instantiated in the context of policy research.
    \item In Study 1, we present an evaluation of the workflow with policy researchers that explores how this workflow fits into existing research practices, how it compares to their own non-AI methods.
    \item In Study 2, we provide a large-scale case study on AI’s economic impacts that applies the workflow to both Reddit discussions and interview transcripts, and compare the resulting themes to authoritative reports, highlighting where the workflow can supplement and extend current policy research methods.
\end{itemize}}

Taken together, we offer empirical evidence about when and how LLM-based thematic analysis can support early-stage policy sensemaking by evaluating the workflow with policy researchers and across multiple data sources, and we surface design and methodological considerations for integrating such tools into policy research workflows.
\section{Related Work}

\subsection{Data Sources of Policy-Relevant Text}

Policy experts have long relied on investigation of public opinion as a source of evidence for policy-relevant decision-making~\cite{manza2002democratic}. To do so, researchers traditionally employ methods such as surveys, opinion polls, listening sessions, and structured interviews, which use standardized instruments and sampling strategies to produce representative estimates of population-level attitudes~\cite{britannica_opinion_poll, berinsky2017measuring, simmons1993capturing}. These methods play a central role in informing policymaking, shaping market strategies, and supporting advocacy efforts~\cite{nancarrow2004polls, rasmussen2018little, burstein2014american}. These methods remain essential when decision makers need insights into the thinking of the wider population. However, they are expensive, slow to field, and constrained by coordination challenges between participants and human interviewers~\cite{hillygus2011evolution}. As a result, they are not always well-suited for rapidly evolving topics or for exploring the full range of how people talk about an issue.

In parallel, the growth of social media and online forums has created new sources of policy-relevant discourse~\cite{reveilhac2022systematic}. Researchers in HCI, CSCW, and computational social science use platforms such as Reddit to analyze people’s perceptions about privacy~\cite{li2021developers}, political acts and policy~\cite{kim2025capturing, dogan2024narrative}, the needs of marginalized groups~\cite{nagaraj2025rideshare}, and responses to new technologies~\cite{knittel2021anyone}. These platforms provide unsolicited narratives, concrete experiences, and sometimes anonymous, critical, or fringe viewpoints that can be hard to elicit in structured surveys or interviews.

However, using online platforms for policy research also raises practical and ethical challenges around data access, consent, and platform governance. Their user bases are imbalanced and non-representative of the general population~\cite{ferrer2021discovering}. Moreover, platforms have tightened access to APIs and bulk data over time, and researchers must respect evolving terms of service and legal constraints~\cite{mimizuka2025post, fiesler2020no}. Scholars have also highlighted the importance of protecting user privacy, avoiding harm to vulnerable communities, and being transparent about data collection and reuse~\cite{vitak2016beyond, fiesler2023internet, fiesler2024remember}. These considerations motivate careful selection of subreddits and posts, as well as anonymization and aggregation when presenting quotes. In our work, we treat Reddit as one deliberately filtered source of policy-relevant discourse that complements other forms of data rather than replacing them. 

Interviews are a valuable way to gather policy-relevant insights alongside demographic information. Recent work has demonstrated the potential of LLM-based chatbots as effective channels for collecting interview data~\cite{wong2025ai, kim2024llm}. Building on this, researchers have developed tools that leverage LLM-based chatbots for large-scale interviews and have shown their utility in qualitative data collection~\cite{geiecke2024conversations}. However, conducting traditional thematic analysis on large volumes of interview data remains time-consuming and resource-intensive. In this work, we therefore treat chatbot-led interviews as a promising data source that provides demographic richness while enabling us to examine how an LLM-assisted thematic analysis workflow can support policy research.

Taken together, these strands of work suggest a spectrum of data sources for policy-relevant text: representative surveys, listening sessions and polls, online forums, and transcripts of interviews. Each source offers different strengths and weaknesses in terms of representativeness, depth, and logistical cost. \textbf{In this work, our goal is not to replace those traditional methods or to validate whether chatbots are good interviewers. Instead, we combine chatbot-led interviews and online forum posts as complementary text sources and focus on how an existing LLM-assisted thematic analysis workflow can support early-stage policy sensemaking.}

\subsection{Computational Methods for Structuring and Analyzing Policy-Relevant Text}

As these diverse channels produce growing volumes of text, researchers have developed computational methods to organize this material into themes. A long line of work uses non-LLM techniques such as topic models and clustering to structure large text corpora into interpretable components~\cite{blei2003latent}. In public opinion and policy-adjacent domains, topic models have helped analyze social media conversations e.g., Reddit discussions of deepfakes and their societal implications~\cite{gamage2022deepfakes}. These methods can reveal broad topical structures and trends, but they often require substantial technical expertise and focus on optimizing statistical properties rather than aligning directly with policy researchers’ workflows.

Beyond static topic models, interactive and dynamic systems help analysts iteratively surface and refine themes in large datasets. Prior work on dynamic surveys, interactive visual analytics, and mixed-initiative tools enables users to explore text corpora, adjust topic granularity, and incorporate domain expertise into the modeling process~\cite{lei2025dynamic, shapiro2025exploratory, chen2025dango}. These systems highlight the value of keeping human analysts in the loop rather than relying on fully automated pipelines. However, they typically target data scientists or visualization experts, and the specific focus on the needs and practices of policy researchers is underexplored.

LLMs have also been used as instruments in social and HCI research, generating synthetic personas or simulated respondents. Recent work evaluates LLM-generated synthetic HCI research data~\cite{hamalainen2023evaluating}, examines how LLMs compose persona descriptions~\cite{salminen2024deus}, and reflects on the challenges and opportunities of LLM-based synthetic personae and data~\cite{prpa2024challenges}. These studies ask whether LLMs can stand in for human participants or help researchers reason about different types of respondents. Our work takes a different stance: we treat LLMs as tools for structuring human-generated text from interviews and forums, rather than as substitutes for people.

Recent work in HCI and qualitative methods explores how LLMs can assist with coding, memoing, and generating candidate themes from interview or forum data~\cite{kang2025themeviz, wang2025lata, kapania2025m, gao2024collabcoder}. Closest to our goal, we have identified LLM-assisted thematic analysis frameworks that use multi-stage prompts to move from unstructured text to topics and themes. Tools such as TopicGPT, DocETL, LLooM and QuaLLM show that LLMs can propose topics, cluster documents, and support exploratory analysis over large corpora~\cite{pham2024topicgpt, lam2024concept, shankar2024docetlagenticqueryrewriting, rao2024quallm}. TopicGPT clusters large corpora into topics and produces natural-language topic descriptions, allowing for quick mapping of what people are talking about at scale~\cite{pham2024topicgpt}. DocETL operationalizes pre-defined analytical constructs to process various formats of data in a repeatable and auditable way~\cite{shankar2024docetlagenticqueryrewriting}. LLooM organizes large collections of unstructured text with interactive, iterative data visualizations~\cite{lam2024concept}.
QuaLLM demonstrates how an LLM can analyze online discussion forums and extract themes through a multi-phase prompting approach, reducing manual coding effort while preserving researcher oversight~\cite{rao2024quallm}. These studies highlight both the promise of faster analysis and concerns about prompt sensitivity, opacity, and hallucinated structure.

Despite these advances, many of the LLM-assisted workflows remain code-centric and prompt-engineering-heavy, which limits their accessibility for policy practitioners. Some systems require users to write scripts, manage APIs, or tune prompts manually; others, such as Reddit Answers and The Giga Brain, expose LLM-powered search over forums but do not give users fine-grained control over data sources or analysis stages\footnote{\url{https://www.reddit.com/answers/}}\footnote{\url{https://thegigabrain.com/feed}}. Tools that provide a workflow that non-programmer policy researchers can adopt, inspect, and critique are underexplored. Moreover, prior work has not evaluated how an LLM-assisted thematic analysis framework fits into real policy research or how its outputs compare to authoritative reports that currently guide policy making.

Taken together, prior research offers strong building blocks for LLM-based computational analysis of policy-relevant unstructured text. However, we still lack evidence on whether an LLM-assisted text analysis framework can help policy researchers make sense of interview transcripts and forum data, or how its outputs compare to authoritative reports. \textbf{In this paper, we extend and apply QuaLLM in a policy context, and evaluate its use through a formative study with policy researchers and a case study on AI’s economic impacts that compares its outputs to existing authoritative policy reports. Design considerations of choosing QuaLLM are illustrated in the next section. We'll also discuss implications and broader generalizations of LLM-assisted frameworks later in this paper.}

\deleted{Recent advances in LLMs have enabled new approaches to analyzing unstructured text, such as online discussions on forums like Reddit. Tools like TopicGPT, DocETL, and LLooM \cite{pham2024topicgpt, lam2024concept, shankar2024docetlagenticqueryrewriting} can extract topics from large text corpora, but they typically require technical expertise in prompt engineering, API integration, and programming. QuaLLM \cite{rao2024quallm} demonstrated the potential for LLMs to analyze online discussion forums and extract themes through a multiphase prompting approach, yet lacks an accessible system for users. Other LLM-wrapper tools such as Reddit Answers \footnote{\url{https://www.reddit.com/answers/}} and The Giga Brain\footnote{\url{https://thegigabrain.com/feed}} have made search more accessible, but don't allow users control over the actual data sources (e.g., which subreddits). \Sys helps overcomes these gaps and enables thematic analysis at scale. 
}

\section{Workflow}
\subsection{Design Goals}
\label{sec:design-goals}
We derived our design goals from two inputs: first, we drew on our team's policy research experience. Second, we drew on federal guidance and law that emphasize building evidence for policymaking and improving how agencies learn from public participation and lived experience \cite{omb_memo, omb_ppce_rfi_2024, Policymaking_Act_18}. Together, these inputs informed the design of an LLM-assisted workflow intended to support early-stage policy sensemaking. Specifically, we articulate the following design goals:
\begin{itemize}
    \item \textbf{Rapid scoping under tight timelines.} The workflow should transform large volumes of unstructured text into a structured, survey-like map of concerns and their relative prevalence, enabling researchers to quickly orient themselves and identify directions for further investigation.
    \item \textbf{Scaling to naturally occurring discussion.} The workflow should ingest and synthesize public discourse at scale, summarizing it into concerns and themes. This provides a complementary evidence stream when traditional methods such as interviews or surveys are infeasible due to time, cost, or access constraints.
    \item \textbf{Source-aware interpretation.} The workflow should make data sources explicit and easy to change, presenting results as reflective of a specific community or corpus rather than the general public.
    \item \textbf{Prevalence-guided prioritization} The workflow should surface which concerns appear most prevalent within the selected source, supporting triage, agenda-setting, and the prioritization of follow-up research or engagement.
    \item \textbf{Human oversight and policy framing.} The workflow should include clear points where researchers define scope, framing prior to reviewing results, ensuring outputs remain aligned with a policy question rather than producing unconstrained or decontextualized topics.
\end{itemize}
Our five goals require a workflow that rapidly converts naturally occurring public discourse into a structured, source-specific map of concerns with prevalence signals, while preserving clear points for human framing and oversight. We build on QuaLLM~\cite{rao2024quallm} because it was designed to extract survey-like themes from unstructured text, estimate prevalence through quote-to-theme mappings, and keep analysis interpretable through explicit intermediate artifacts. It is also modular and easily adaptable to specific uses. To justify this choice, we also explicitly compare QuaLLM to TopicGPT\cite{pham2024topicgpt} and LLooM\cite{lam2024concept}—two closely related LLM-assisted frameworks that also make sense of large-scale unstructured text. Table~\ref{tab:framework-comparison} shows how TopicGPT, LLooM, and QuaLLM align with our five design criteria, and why QuaLLM best fits our needs. We therefore use QuaLLM as the thematic analysis component of our workflow.

\begin{table*}[tb]
\centering
\footnotesize
\setlength{\tabcolsep}{6pt}
\renewcommand{\arraystretch}{1.2}
\begin{tabularx}{\textwidth}{p{0.15\textwidth} X X X}
\toprule
\textbf{Design Goals (Section \ref{sec:design-goals})} & \textbf{TopicGPT~\cite{pham2024topicgpt}} & \textbf{LLooM~\cite{lam2024concept}} & \textbf{QuaLLM~\cite{rao2024quallm}} \\
\midrule

\textbf{Rapid scoping under tight timelines} &
\textit{Partial.} Efficient for generating and assigning topics, but translating them to themes and sub-themes still requires additional human interpretation time. &
\textit{Partial.} Runs quickly, but producing a policy-ready concern summary can require repeated human refinement of concepts across several iterations. &
\textbf{Strong.} Produces a quick draft of prioritized list of themes and sub-themes.\\

\midrule
\textbf{Scaling to naturally occurring discussion} &
\textit{Partial.} Scales via sample-based topic generation plus corpus-wide assignment, but is geared toward topic discovery rather than end-to-end concerns synthesis. &
\textbf{Strong.} Built for large-scale text: distills evidence into concepts and applies/scores them across the dataset. &
\textbf{Strong.} Demonstrated at scale (e.g., $>$1M comments) and explicitly supports extracting, aggregating, and quantifying concerns. \\

\midrule
\textbf{Source-aware interpretation} &
\textit{Partial.} Can run on any corpus, but does not explicitly frame results as community/source-specific in the output structure. &
\textbf{Strong.} Compares concept prevalence across user-defined slices (e.g., source/subgroup) using metadata. &
\textbf{Strong.} Frames outputs as tied to the selected forum/community, not population-representative. \\

\midrule
\textbf{Prevalence-guided prioritization} &
\textit{Partial.} Can filter rare topics and report frequencies, but does not foreground ranked concerns for triage. &
\textbf{Strong.} Makes prioritization explicit via prevalence visualizations across slices (what matters most, where). &
\textbf{Strong.} Produces prevalence-ranked sub-themes from aggregated quote-level evidence. \\

\midrule
\textbf{Human oversight and policy framing} &
\textit{Partial.} Allows steering (examples/edits), but offers fewer explicit checkpoints for pre-specifying framing and auditing intermediates. &
\textbf{Strong.} Keeps humans in the loop via editable concepts and explicit inclusion criteria (add/edit/merge/split). &
\textbf{Strong.} Requires human-defined framing and exposes inspectable intermediate artifacts that constrain synthesis to the policy question. \\

\bottomrule
\end{tabularx}
\caption{Comparison of TopicGPT, LLooM, and QuaLLM against our Section \ref{sec:design-goals} design goals. QuaLLM best matches our goal of producing a source-specific, prevalence-aware, survey-like map of concerns while preserving explicit points of human framing and oversight.}
\label{tab:framework-comparison}
\end{table*}

\subsection{Workflow Overview}
\begin{figure}[hbt!]
    \centering
    \includegraphics[width=\linewidth]{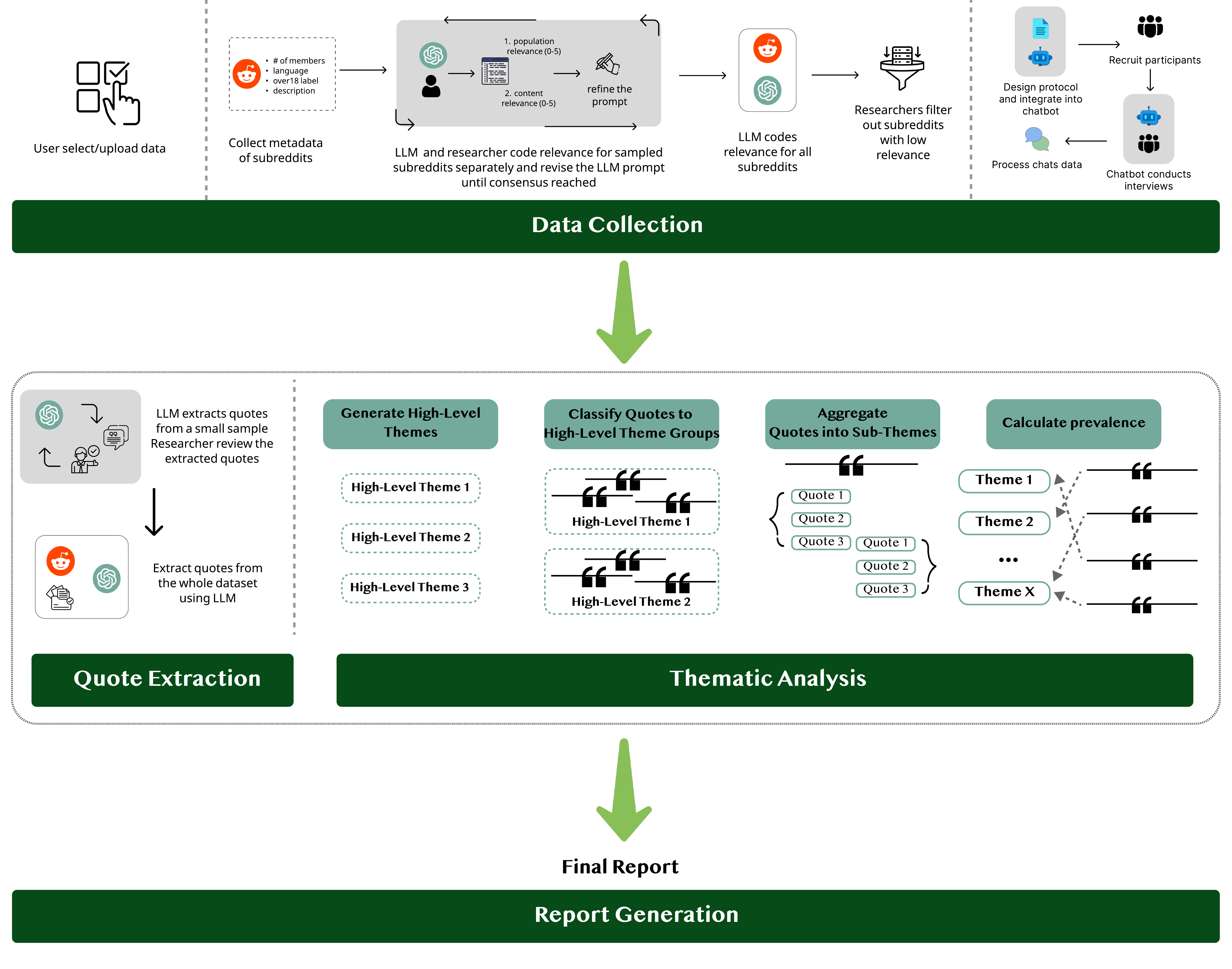}
    \caption{Overview of the Workflow. The workflow comprises four stages, where the two stages(Quote Extraction and Thematic Analysis) are adapted from an existing LLM-based thematic analysis workflow called QuaLLM. We demonstrated details of each stage in Sections 3.1, 3.2, 3.3, and 3.4 respectively.}
    \Description{Overview of the Workflow. The workflow comprises four stages, where the two stages(Quote Extraction and Thematic Analysis) are adapted from an existing LLM-based thematic analysis workflow called QuaLLM. We demonstrated details of each stage in Sections 3.1, 3.2, 3.3, and 3.4 respectively.}
    \label{fig:workflow}
\end{figure}

In this work, we develop a workflow that adapts an existing LLM-based thematic analysis workflow(QuaLLM) for application in policy research~\cite{rao2024quallm}. The workflow comprises four stages: data collection, quote extraction, thematic analysis, and report generation. The fist stage involves selecting appropriate data sources for the research topic. The quote extraction stage focuses on extracting topic-relevant quotes from given data sources. The thematic analysis stage derives themes from these unstructured texts and ranks the representative ones after mapping each quote to its corresponding theme~\cite{rao2024quallm}. Finally, the workflow generates a report to provide policy researchers with an interpretable synthesis of the underlying data. Figure ~\ref{fig:workflow} shows the overview of the workflow.

\subsubsection{Data Collection}
Data collection involves selecting appropriate data sources for a given policy topic and gathering relevant content from those sources. This process can be adjusted based on constraints such as time, budget, and available resources. In this work, we present three example data collection approaches that vary in scale and degree of human involvement. In Study 1, we pre-downloaded data from multiple subreddits on Reddit and asked participants to manually select one subreddit. In Study 2, we designed semi-automated approaches in which an LLM collects data based on predefined instructions, with human involved for tuning. Further details of each data collection method are described in the corresponding study sections. 

\subsubsection{Quote Extraction}
In this section, we demonstrate the procedure of quote extraction from online forums using a Reddit dataset.

\paragraph{The Reddit Dataset}
Due to its vast user base, Reddit is a valuable source for gauging public sentiment on policy-relevant topics. Its anonymity-driven candid discussions and specialized subreddit communities facilitate rich contextual data and longitudinal analysis of public opinion \cite{fiesler2024remember, nagaraj2025rideshare, chen2021using, xu2024public, huang2024politically}.  Furthermore, comprehensive data is easily and publicly accessible. We used Reddit data from The Eye\footnote{\url{https://the-eye.eu/redarcs/}} Reddit archive (Study 1) and Academic Torrents\footnote{\url{https://academictorrents.com/}} (Study 2), which contains submissions (the original posts that initiate threads in subreddits, consisting of a title with optional text, link, or media) and comments (responses to submissions or other comments that form threaded discussions). In the studies presented in this work, we downloaded and selected subreddits related to our research topics, with the specific selection procedures described in detail in Sections 4 and 5, respectively. 

\begin{figure}
    \centering
    \includegraphics[width=0.7\linewidth]{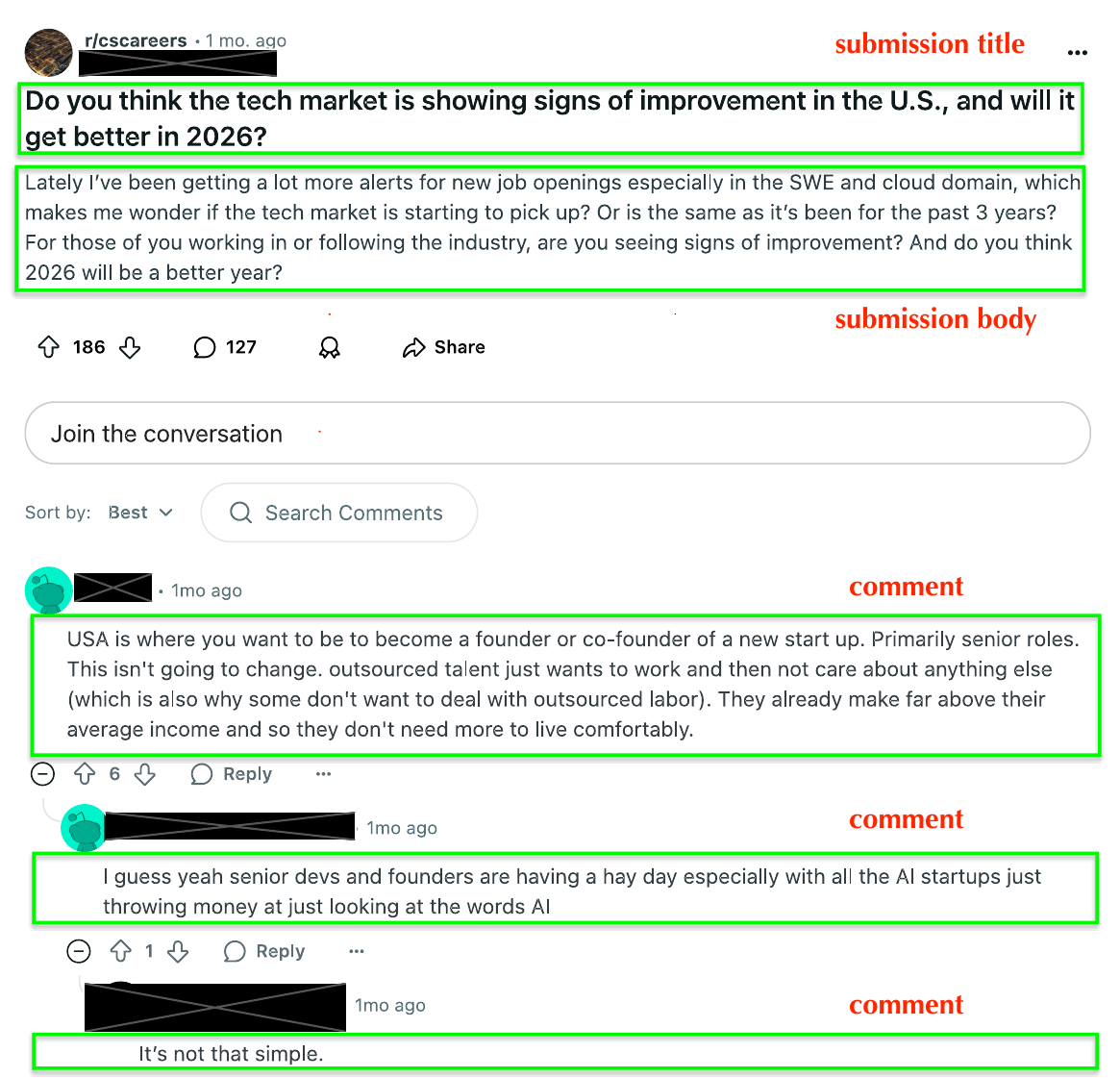}
    \caption{Example of a discussion in r/cscareers subreddit}
    \Description{Example of a discussion in r/cscareers subreddit. From top to down, there is a submission title, a submission body, and three comments.}
    \label{fig:submission_comments}
\end{figure}

\paragraph{Data Processing and Quote Extraction from Reddit}
We organized the downloaded subreddit files into structured JSON formats. Each entry represented a discussion with four fields: submission ID, submission title, submission body, and a sequential list of comments. See the annotation in Figure \ref{fig:submission_comments} for an example. We stored comments in linear order without nesting, which sacrificed some relational context but made processing more efficient. We then fed each entry into the LLM with instructions to extract quotes relevant to the research topic. For every extracted quote, the output JSON includes the submission ID, submission title, and a summary of the submission body to provide context. To validate the LLM’s performance and reduce hallucination~\cite{huang2023survey}, we randomly sampled 10 subreddits and 100 entries from each (1,000 discussions in total) in each iteration. The LLM extracted relevant quotes from these samples, and one researcher manually reviewed the results to check that each quote matched the original sentence. The researcher kept randomly sampling and refining the prompt through multiple iterations until the extractions reached over 70\% agreement on relevance. Details of the LLM prompts used in this stage are attached in Appendix ~\ref{appendix:prompting} 

\subsubsection{Thematic Analysis}
Following the QuaLLM ~\cite{rao2024quallm}, our workflow started with defining high-level themes of the selected research topic. Then we used LLM to categorize quotes into each high-level theme group. Within each group of quotes, we then leveraged LLM to extract themes in a bottom-up manner. Next, we instructed the LLM to map each quote to its appropriate theme. Within each high-level theme group, we count the number of quotes for each theme and find the representative themes. Emphasizing a stepwise policy research workflow, each prompt is engineered to support policy-focused analytical objectives. We detail our prompt engineering approach for each step in Appendix~\ref{appendix:prompting}.

\subsubsection{Report Generation}
In this stage, we generate a report with high-level themes, description of themes (generated by LLM), and representative quotes. To validate the integrity of the outputs, in this work, two researchers manually reviewed every theme in every report, along with three randomly sampled quotes mapped to it to confirm that the quotes were drawn from the original dataset and were not fabricated by the LLM. 
\section{Study 1: An Evaluation of the Workflow With Policy Researchers Using Limited Online Forum Discussions}
\begin{figure}[htb!]
  \centering
  \includegraphics[width=0.7\textwidth]{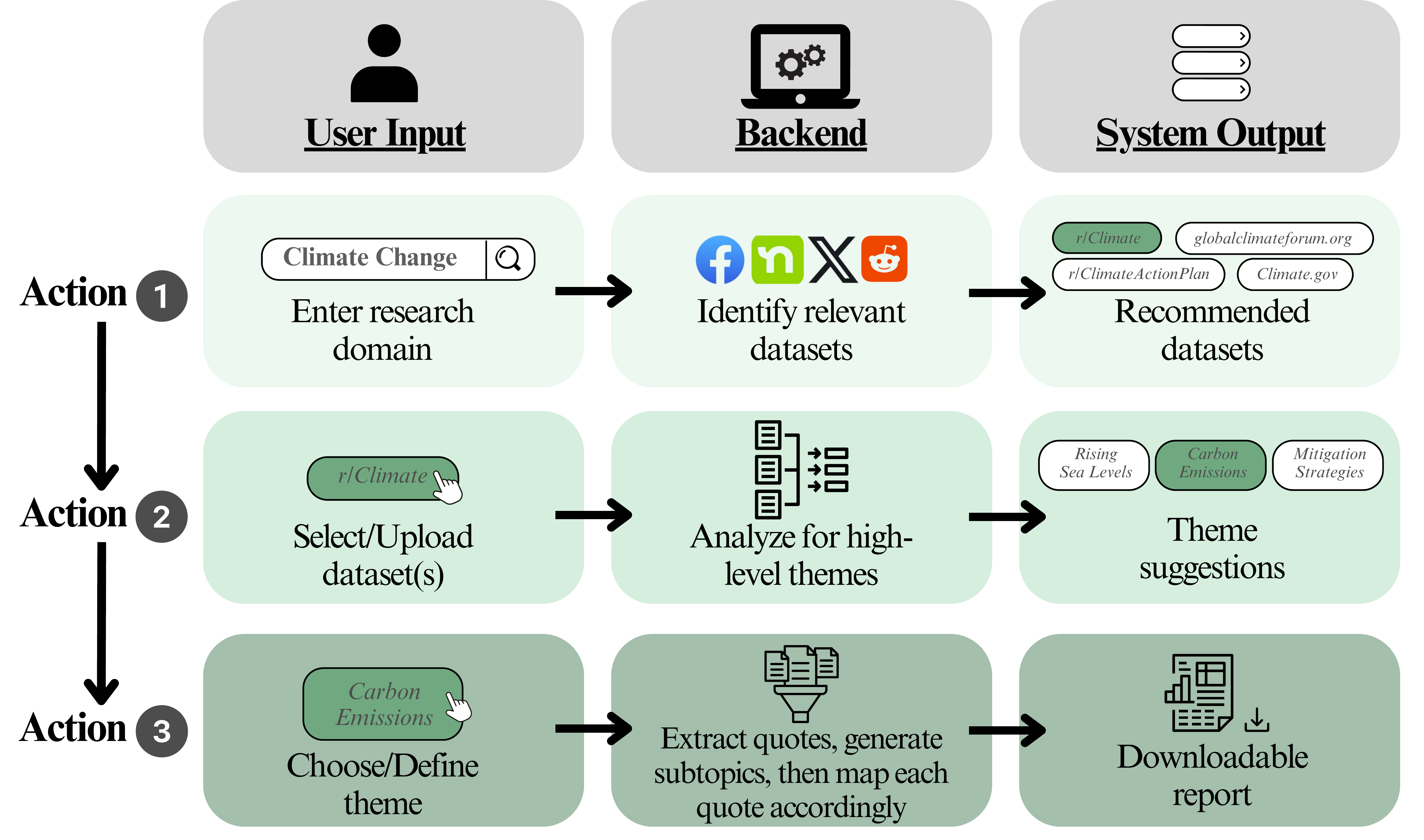} 
  \caption{Illustration of How User Interact with the Workflow in This Study. Action 1: The user enters a policy-relevant topic (left), prompting the backend (center) to identify relevant datasets, which the system (right) then presents as recommendations. Action 2: The user selects or uploads the dataset(s), triggering the backend to analyze for high‐level themes, and the system outputs theme suggestions. Action 3: The user chooses or defines a final theme, prompting the backend to extract relevant quotes, generate subtopics, and map each quote accordingly, resulting in a downloadable report.}
  \Description{Illustration of How User Interact with the Workflow in This Study. Action 1: The user enters a policy-relevant topic (left), prompting the backend (center) to identify relevant datasets, which the system (right) then presents as recommendations. Action 2: The user selects or uploads the dataset(s), triggering the backend to analyze for high‐level themes, and the system outputs theme suggestions. Action 3: The user chooses or defines a final theme, prompting the backend to extract relevant quotes, generate subtopics, and map each quote accordingly, resulting in a downloadable report.}
  \label{fig:SystemArch}
\end{figure}

To evaluate how LLM-assisted tools can support policy researchers, we implemented a user interface on top of the workflow (see Figure \ref{fig:SystemArch} for a high level overview of the user interaction) and conducted a user study with N=11 policy researchers. We now discuss the methods and findings from the user study.

\subsection{User Interface Implementation}

\begin{figure}[hbt!]
    \centering
    \includegraphics[width=0.8\textwidth]{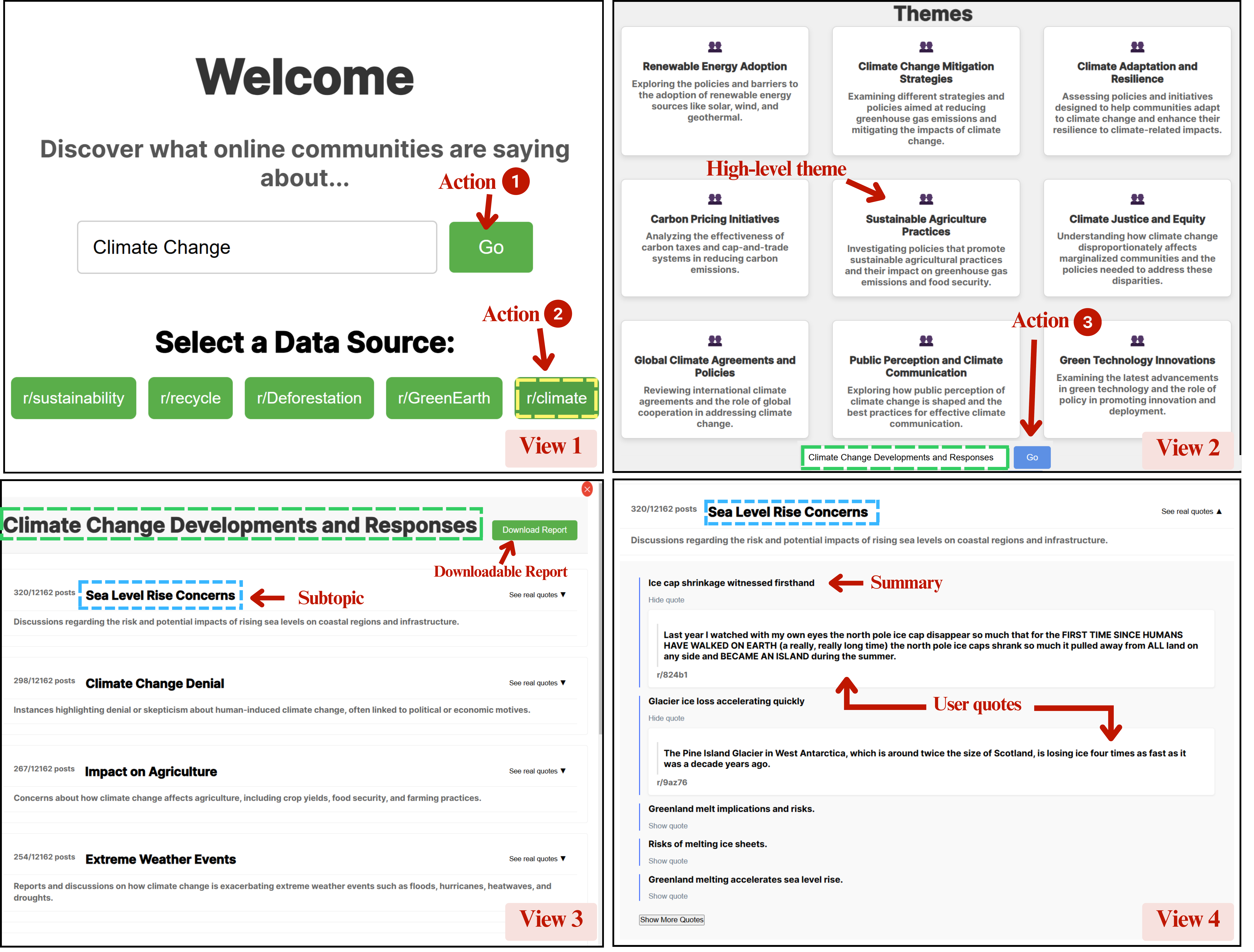}
    \caption{\added{Screenshots of the User Interface in this study. View 1 shows the interface. This view prompts users to input a research domain for analysis (Action 1) and select a data source (Action 2). View 2 shows high-level themes in the workflow interface. The user can then select or search for a primary research topic (Action 3). View 3 shows the final Report View of the interface, displaying all subtopics identified from the data source and quote counts. View 4 shows examples of 5-6 word summaries beneath the subtopic and a couple of fully displayed quotes on which the summaries are based.}}
    \Description{Screenshots of the User Interface in this study. View 1 shows the interface. This view prompts users to input a research domain for analysis (Action 1) and select a data source (Action 2). View 2 shows high-level themes in the workflow interface. The user can then select or search for a primary research topic (Action 3). View 3 shows the final Report View of the interface, displaying all subtopics identified from the data source and quote counts. View 4 shows examples of 5-6 word summaries beneath the subtopic and a couple of fully displayed quotes on which the summaries are based.}
    \label{fig:Screens}
\end{figure}

\subsubsection{Data Source Selection}
Users specify a policy domain (e.g.,  ``Climate Change'') and select relevant data sources. The system uses an LLM to match topics with policy-relevant online communities, currently providing options from subreddit forums but designed for expansion to other platforms and alternative data sources (Figure~\ref{fig:Screens}, View 1).
\subsubsection{Theme Generation}
The workflow identifies high-level themes within the selected data source. Users can explore suggested themes or perform custom searches based on their own research needs (Figure~\ref{fig:Screens}, View 2).
\subsubsection{Report Generation}
For each selected or entered theme, the workflow processes data through a series of prompts. This multi-stage pipeline ensures structured, concise insights tailored for policy research. Specifically:
\begin{itemize}
    \item Relevant quotes of people's actual experiences and anecdotes are extracted using a Quote Extraction prompt designed to minimize bias.
    \item Aggregated quotes are analyzed to identify subtopics (Figure~\ref{fig:Screens}: View 3).
    \item Quotes are mapped to appropriate subtopics to ensure structured organization.
    \item Concise summaries (5–6 words) are generated for enhanced readability (Figure~\ref{fig:Screens}: View 4).
    \item A final downloadable report is created for offline analysis.
\end{itemize}

The system processes approximately 1,000 quotes per 10 minutes, caching results for immediate access on subsequent views (detailed prompts in Appendix~\ref{appendix:prompting}).
\subsubsection{Implementation Details}
We used RESTful APIs to orchestrate interactions between the frontend user interface and the workflow. Figure ~\ref{fig:Screens} shows the screenshots of the user interfaces. The core functionality of the implementation includes data preprocessing through Pandas, caching for iterative analysis, and JSONL-based storage for structured reports. While currently leveraging The Eye Reddit archive, the modular architecture supports structured processing of any user-uploaded dataset. Furthermore, this allows the user agency over data sources, which current LLM-wrapper tools such as Reddit Answers\footnote{\url{https://www.reddit.com/answers/}} and The Giga Brain\footnote{\url{https://thegigabrain.com/feed}} do not allow. This is especially important for policy researchers looking to explore public sentiment using only specific data sources they deem to satisfy their requirements. 

The initial script that handles downloaded data scraped from The Eye aggregates posts and their associated comments into unified discussion threads to preserve contextual coherence, ensuring that analyses accurately reflect the temporal order of public discourse. The output raw data is stored in CSV format to optimize access and preprocessing efficiency. The Flask backend processes this raw data with the workflow described in the previous section.

\subsection{Method}
\begin{figure}[htb]
    \centering
    \includegraphics[width=0.8\linewidth]{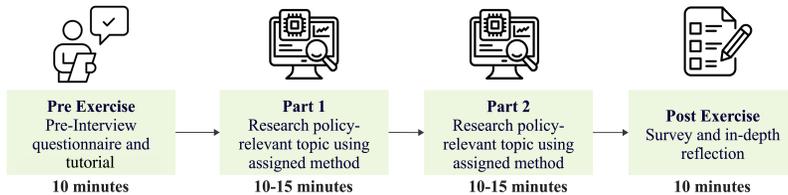}
    \caption{\added{Timeline of the User Study. This figure illustrates the four-phase research method comparing AI-assisted and participants' own non-AI expert approach. The process begins with an interview and tutorial, followed by two sequential research parts randomized based on topic (Climate Change or Social Media \& Kids) and method (with the workflow or own non-AI expert approach), and concludes with a retrospective survey and interview. Each research phase is allocated 10-15 minutes, with participants documenting both quantitative and qualitative findings using standardized interviewee report templates throughout the process.}}
    \Description{Timeline of the User Study. This figure illustrates the four-phase research method comparing AI-assisted and participants' own non-AI expert approach. The process begins with an interview and tutorial, followed by two sequential research parts randomized based on topic (Climate Change or Social Media \& Kids) and method (with the workflow or own non-AI expert approach), and concludes with a retrospective survey and interview. Each research phase is allocated 10-15 minutes, with participants documenting both quantitative and qualitative findings using standardized interviewee report templates throughout the process.}
    \label{fig:timeline}
\end{figure}

\begin{table}[h!]
\centering
\begin{tabular}{p{0.5cm} p{6.8cm} p{4.7cm} p{1.3cm}}
\hline
\textbf{PID} & \textbf{Job and Position} & \textbf{Expertise} & \textbf{Education} \\
\hline
1 & Deputy Director of Policy \& Strategy at a pediatric research organization & Health policy & MS \\
2 & Research Director at a university public policy center& Media and social influences & PhD \\
3 & Postdoctoral Research Associate at education/public affairs institutes & Education policy & PhD \\
4 & Strategic management lead at a philanthropic policy organization & Medicaid policy & MS \\
5 & Social impact professional and federal agency fellow in energy/international affairs & Global anti-poverty policy & MS \\
6 & Lead Economist at a national central banking institution & Public finance policy & MS \\
7 & Professor at a public affairs and administration school & Public policy and administration & PhD \\
8 & Deputy Chief of Staff for Economic Growth in a U.S. state government & State economic policy & MS \\
9 & Policy Advisor in a U.S. state governor’s office & Health and Human Services policy & MS \\
10 & Former federal advisory board policy analyst and communications officer & Domestic policy & MS \\
11 & University Professor; former Dean of a public affairs school & Public budgeting and finance & PhD \\
\hline
\end{tabular}
\caption{\added{Job, Position, Expertise and Education of the Policy Researchers in the Interview}}
\Description{Job, Position, Expertise and Education of the Policy Researchers in the Interview}
\label{tab:policy_experts}
\end{table}

Figure \ref{fig:timeline} shows an overview of the study procedure. We recruited 11 experienced policy researchers to evaluate the workflow through comparative analysis with participants' own non-AI expert approaches. Basic information of the participants is shown in Table ~\ref{tab:policy_experts}. During 45-60-minute interviews, participants first completed a pre-task survey assessing their policy research experience, followed by a structured comparison task. Participants were randomly divided into two groups based on the order of topics they would research. Within each group, they were further divided based on the order of methods used (workflow vs. participants' own non-AI expert approach). This organizational structure helped to control for order effects, such as practice or fatigue, which could influence participants' performance depending on the sequence of tasks. Each participant spent 10-15 minutes researching one of two randomly assigned topics using one method and then switched to the other method for the second topic. For both topics, they recorded themes and anecdotal evidence in standardized worksheets (see Appendix ~\ref{sec:Interview Worksheet}), enabling a direct comparison of research efficiency and effectiveness.

We audio-recorded and transcribed all interviews, then conducted thematic analysis using open coding. We coded the transcripts to identify recurring themes and patterns in user feedback. We iteratively refined these codes through discussion until reaching consensus, then grouped them into higher-level themes. This analysis revealed three primary themes: enhancement of traditional methods, interface design benefits, and areas for improvement.

Post-task Likert-scale surveys measured user experience and perceived benefits across multiple dimensions, including speed, breadth of perspectives, anecdote discovery, data quality, and ease of analysis. Participants also reflected on how the workflow compares to and may complement surveys, interviews, and listening sessions. We quantitatively analyzed the worksheet data to compare the number of themes gathered within the time constraint across both methods. The study was approved by our institution's IRB.

\subsection{Findings}

\begin{figure}[htb]
  \centering
  \includegraphics[width=0.6\textwidth]{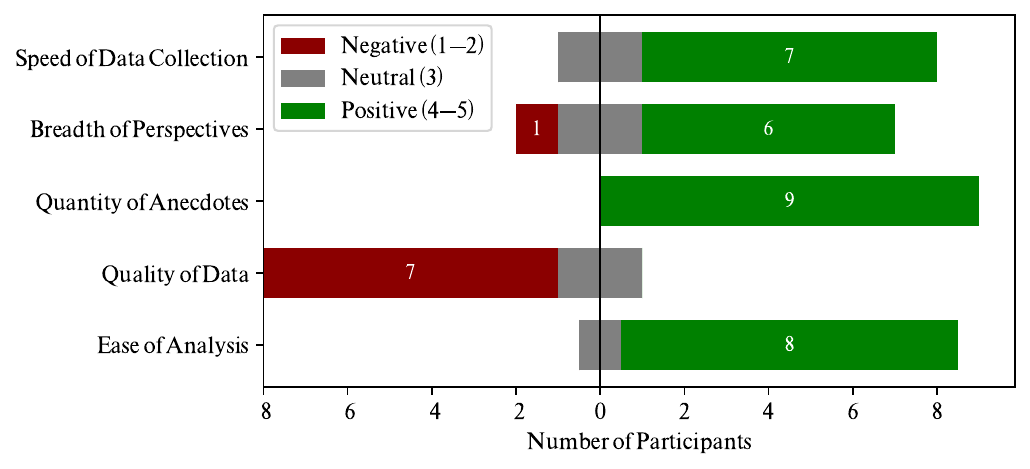}  
  \caption{This diverging stacked bar chart breaks down Likert scale responses (1-5) into three groups (negative, neutral, positive) with the y-axis representing one question from our survey. The length of each colored segment indicates how many of our evaluation participants chose that group. The further left the red bar extends, the more negative; the further right the green bar extends, the more positive. The gray area indicates the number of neutral participants. Notably, most participants found the quantity of anecdotes and ease of analysis provided by the workflow a key strength, but the data quality a major place for improvement.}
  \Description{This diverging stacked bar chart breaks down Likert scale responses (1-5) into three groups (negative, neutral, positive) with the y-axis representing one question from our survey. The length of each colored segment indicates how many of our evaluation participants chose that group. The further left the red bar extends, the more negative; the further right the green bar extends, the more positive. The gray area indicates the number of neutral participants. Notably, most participants found the quantity of anecdotes and ease of analysis provided by the workflow a key strength, but the data quality a major place for improvement.}
  \label{fig:sentimentchart}
\end{figure}



\subsubsection{The Workflow Enhances Traditional Research Methods While Reducing Resource Requirements}

\begin{figure}[htb]
  \centering
  \includegraphics[width=0.3\columnwidth]{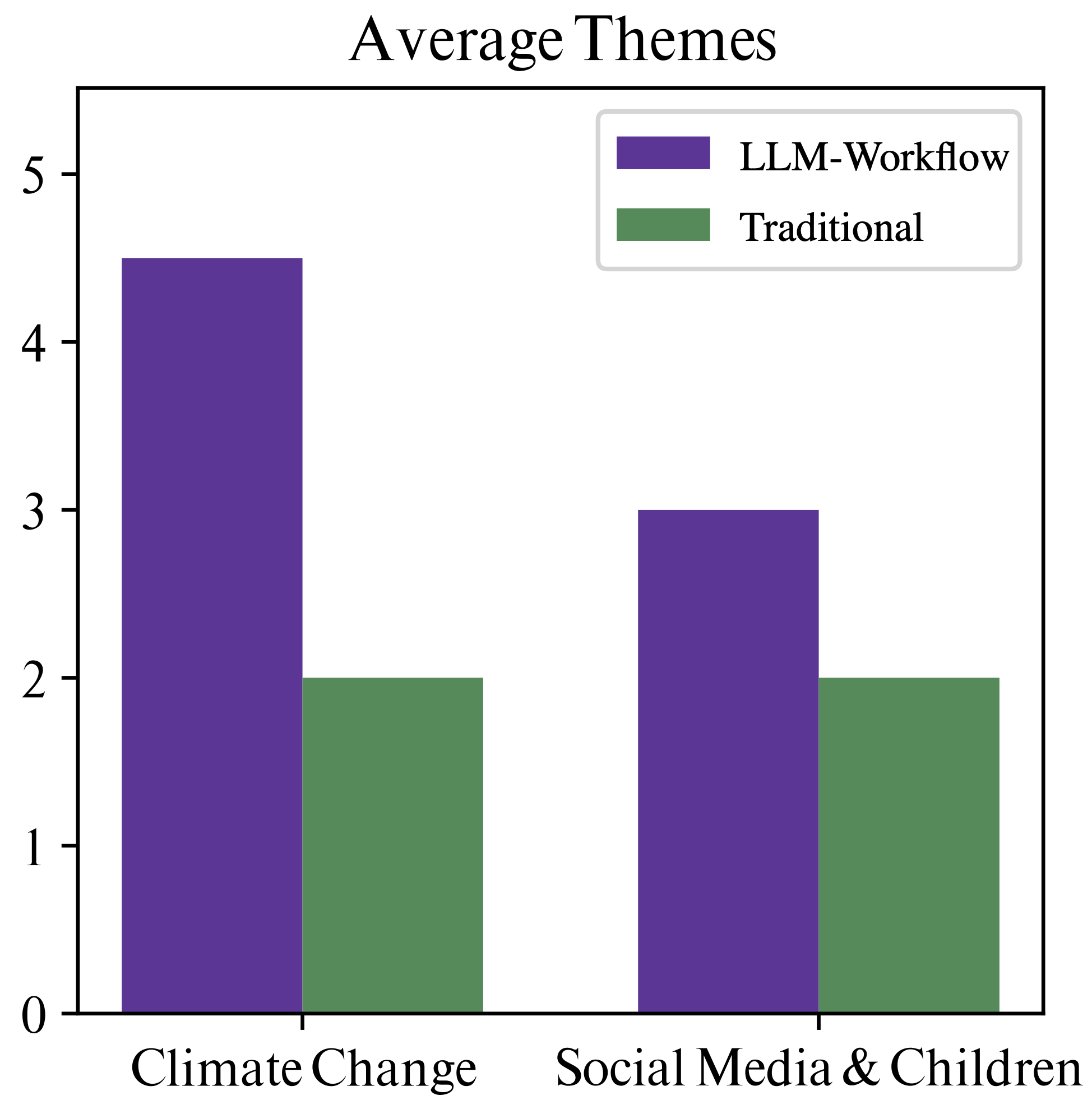}  
  \caption{\added{This bar chart shows the average number of themes gathered across the two research topics---Climate Change and Social Media \& Kids---using two research methods, the workflow and non-AI expert approach. As we can see, even in a limited study duration, using the workflow, on average, allowed participants to collect a higher number of themes for both topics, pointing to the process being an order of magnitude faster.}}
  \Description{This bar chart shows the average number of themes gathered across the two research topics---Climate Change and Social Media \& Kids---using two research methods, the workflow and non-AI expert approach. As we can see, even in a limited study duration, using the workflow, on average, allowed participants to collect a higher number of themes for both topics, pointing to the process being an order of magnitude faster.}
  \label{fig:bargraph}
\end{figure}

The workflow complements existing policy research methods by providing a cost-effective and time-efficient solution. Figure \ref{fig:sentimentchart} demonstrates that 7/11 (64\%) of participants rated workflow's speed and efficiency positively. Furthermore, participants on average collected 2 more themes during the study period compared to their own non-AI expert approach as seen in Figure \ref{fig:bargraph}. With regards to costs, participants told us that listening sessions and surveys cost between \$4,000-\$80,000 respectively, while workflow can operate at much more reduced cost(less than \$1,000 using Azure OpenAI API). Beyond financial benefits, P2 also told us the tool accelerates the public opinion gathering stage, which comprises 10\% of the policy research process: \textit{``Your tool is definitely faster than what we could produce in a survey because we would be surveying hundreds of people'' (P2)}.

The workflow also excels at capturing a broader range of public opinions compared to traditional methods. While conventional approaches exhibit inherent biases toward participants with time and resources, the workflow leverages online forums to expand demographic representation. One participant highlighted this advantage, stating \textit{``Most issues are not very researched in terms of polls and surveys. So this tool could become more versatile'' (P6)}. However, two users noted potential population bias favoring younger, internet-savvy generations, supporting the workflow's role as a supplementary rather than a replacement tool.

All participants believe the workflow would integrate effectively into the existing research pipelines, particularly in early stages. Specifically, they explained the typical policy research pipeline, which starts with background research, consulting existing surveys from think tanks like Pew Research or other governmental agencies and databases, engaging subject matter experts, and finally conducting new surveys or listening sessions -- a process spanning 3-4 months. Participants felt that our tool proved especially valuable for researchers less familiar with topics, as its thematic layout facilitates rapid understanding. As one participant commented, \textit{``I would use your tool to jump start the research process'' (P8).}

\subsubsection{AI Integration and Interface Design Facilitate Objective Data Analysis}
Despite initial AI skepticism, participants found that the workflow's AI backend actually reduces bias in data presentation. Specifically, they felt the thematic categorization limits selection bias, particularly benefiting researchers already familiar with topics. And, the interface effectively presents qualitative data, making non-statistical information more accessible. One participant emphasized this benefit: \textit{``The surveys would say X percent of parents think that students shouldn't use social media. But then you wouldn't get more of the reasoning behind. And I've struggled with this in my previous position in policy research'' (P10)}.

The raw user data, presented as Reddit quotes, distinguishes workflow from non-AI expert research. One participant noted, \textit{``So you know the sense I'm getting is that your tool [is] actually giving us...a little insight into what people are actually saying, as opposed to what people are summarizing about it'' (P2)}. The interface's efficiency impressed participants, with one stating, \textit{``I didn't have to read a bunch of stuff that I wasn't interested in reading'' (P11)}.

The tool also excels at revealing unexpected insights compared to participants' own non-AI expert methods. As one participant observed, \textit{``For unexpected themes I think it's much easier with your tool than with traditional methods'' (P7)}. They also felt that the workflow facilitates access to otherwise hard-to-reach data and anecdotes, enhancing both accessibility and perspective diversity. The Likert scale responses in Figure \ref{fig:sentimentchart} validate these qualitative observations, with 8/11 (73\%) of participants rating the workflow's ease of analysis positively. Additionally, participants consistently gathered more themes using tue workflow across both research topics (Figure \ref{fig:bargraph}), supporting their feedback about improved access to diverse data.

\subsubsection{Areas for Enhancement Focus on Metadata and Trust Building}
Participants identified several opportunities for improvement, primarily centered around metadata enhancement. These concerns are reflected in the quantitative data, where data quality received the lowest positive ratings among all measured dimensions in Figure \ref{fig:sentimentchart}. Diving deeper through retrospective interviews, participants desired demographic information, including geographic, racial, and gender breakdowns, to assess dataset representation. As one participant suggested, \textit{``Do try to categorize Reddit users into some of those demographic data...that's the kind of thing that I would be really interested in'' (P4)}.

The interface requires refinement in data presentation. Participants requested more intuitive quantitative representations, with one noting, \textit{``Sometimes it's helpful to have kind of the quantitative score of how that theme ranks, or what percentage'' (P7)}. Additional suggestions include improved scrolling functionality, reduced quote redundancy, and clearer theme hierarchy labeling.

Finally, two fundamental challenges emerged: inherent AI distrust and limited solution-oriented content. One participant noted, \textit{``People using it for professional research are not necessarily very trusting of it, myself included'' (P4)} with ``it'' referring to AI. Participants suggested linking original Reddit posts to build credibility. The scarcity of actionable solutions stems from the nature of informal forum discussions, which rarely propose concrete policy measures.

\section{Study 2: An Evaluation of the Workflow Against Authoritative Reports Using Large-Scale Online Forum Discussions and Interview Transcripts}
In Study 1, policy researchers found the workflow fast and useful, but they raised three concrete concerns: 1) they lacked demographic and other metadata needed to understand who the Reddit quotes were coming from; 2) they worried that a small, hand-picked set of subreddits would overrepresent younger, internet-savvy users; and 3) they remained cautious about trusting AI-generated summaries without stronger ways to check or validate them. 

\begin{figure}[htb]
    \centering
    \includegraphics[width=\linewidth]{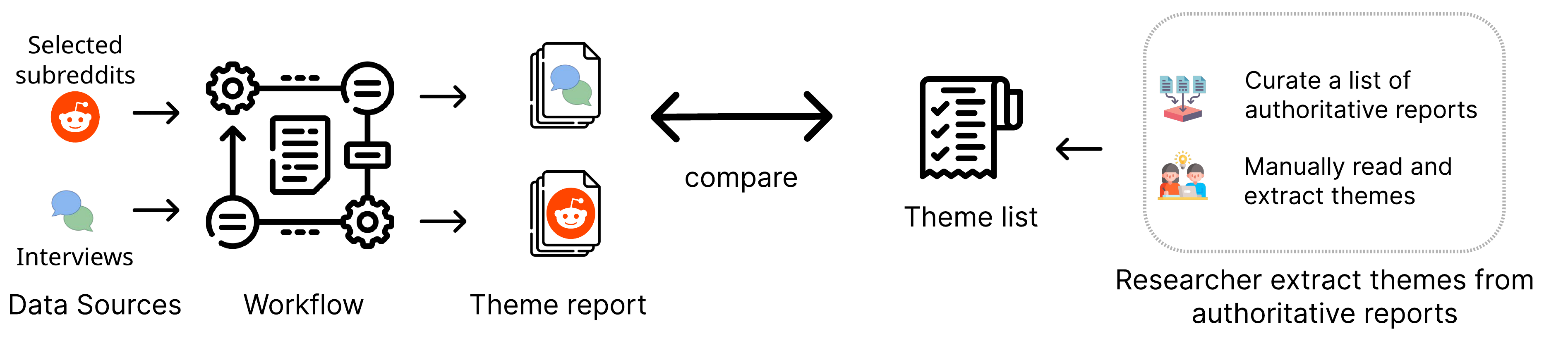}
    \caption{Overview of the evaluation in Study 2. We extend the QuaLLM-assisted thematic analysis workflow to two inputs: (1) Reddit discussions and (2) Chatbot-led interviews with U.S. adults. The workflow applies the same LLM-assisted pipeline to Reddit and interview data to generate theme reports, while a researcher manually reads the policy reports to extract a reference list of themes. We then compare the LLM-generated theme lists and reports against this reference set to evaluate the coverage and alignment of the workflow’s outputs.}
    \Description{Overview of the evaluation in Study 2. We extend the QuaLLM-assisted thematic analysis workflow to two inputs: (1) Reddit discussions and (2) Chatbot-led interviews with U.S. adults. The workflow applies the same LLM-assisted pipeline to Reddit and interview data to generate theme reports, while a researcher manually reads the policy reports to extract a reference list of themes. We then compare the LLM-generated theme lists and reports against this reference set to evaluate the coverage and alignment of the workflow’s outputs.}
    \label{fig:method_overview}
\end{figure}

In Study 2, we keep the same LLM-assisted workflow but extend our evaluation to address these points: 1) acknowledging the limitation of anonymity in Reddit data, we apply the workflow to data collected using traditional methods (interviews) with full demographic information, showcasing its ability to process heterogeneous data formats and contexts. This shows that the workflow remains useful when researchers prefer or require traditional data sources. Specifically, we add 1,058 semi-structured, chatbot-led interviews with U.S. adults for whom we obtained full demographic information. 2) We scale from a handful of forums to a systematically selected set of subreddits drawn from the 40,000 largest communities using subreddit-level metadata and LLM-assisted screening. 3) We assess the reliability and coverage of the workflow’s outputs by comparing its generated themes against themes manually extracted from six authoritative reports on AI and the economy, using these reports as an external reference point. Figure~\ref{fig:method_overview} provides an overview of the evaluation.

\subsection{Methods}

\subsubsection{Data Source: Reddit}
\begin{figure}[htb!]
    \centering
    \includegraphics[width=\linewidth]{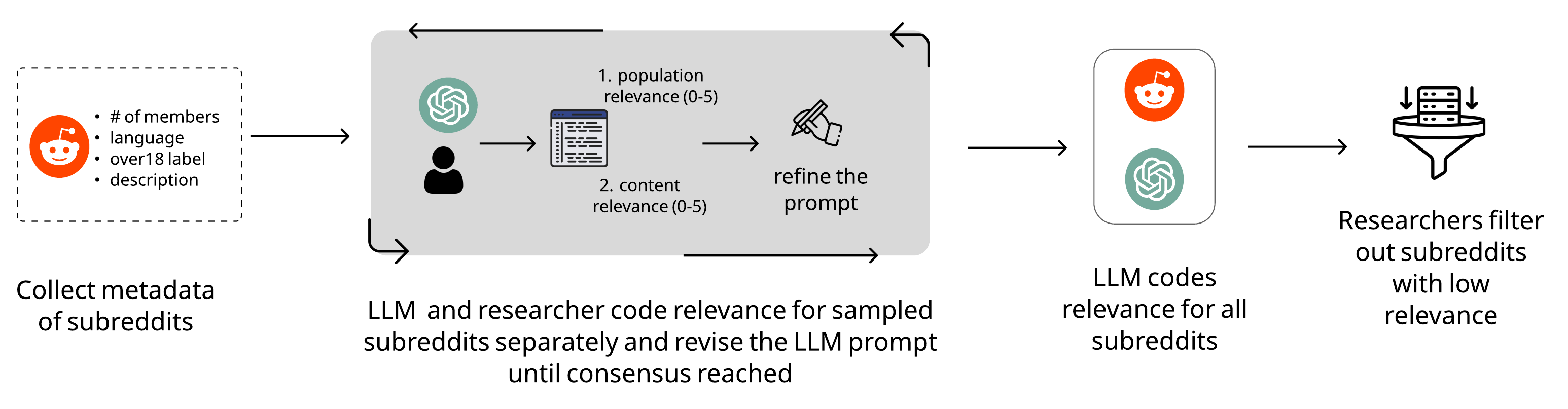}
    \caption{Workflow for selecting and preparing Reddit data in Study 2. Starting from a corpus covering submissions and comments from the top 40,000 subreddits, we collect subreddit-level metadata via the Reddit API (number of members, primary language, adult-content flag, and moderator-written description). A researcher then rates a sample of subreddits on population and topic relevance, and we iteratively refine an LLM prompt so that the model assigns similar 1–5 relevance scores. Using these LLM-assigned scores, we filter out non-English, adult-only, niche, and low-relevance subreddits, retaining a final set of 288 subreddits whose discussions are passed into the quote-extraction and thematic analysis workflow.}
    \Description{Workflow for selecting and preparing Reddit data in Study 2. Starting from a corpus covering submissions and comments from the top 40,000 subreddits, we collect subreddit-level metadata via the Reddit API (number of members, primary language, adult-content flag, and moderator-written description). A researcher then rates a sample of subreddits on population and topic relevance, and we iteratively refine an LLM prompt so that the model assigns similar 1–5 relevance scores. Using these LLM-assigned scores, we filter out non-English, adult-only, niche, and low-relevance subreddits, retaining a final set of 288 subreddits whose discussions are passed into the quote-extraction and thematic analysis workflow.}
    \label{fig:collection-reddit}
\end{figure}

\paragraph{Data Preparation}\deleted{ We used Reddit as an example of online forums in our system, and to be consistent with the interview, we use the topic \dquote{how people in the US perceive the economic impact of AI.} as the main research question. our starting point was an open dataset from Academic Torrents\footnote{https://academictorrents.com/}, which contains submissions (the original posts that initiate threads in subreddits, consisting of a title with optional text, link, or media) and comments (responses to submissions or other comments that form threaded discussions)} \added{In this study, we used a large open Reddit dataset} from Academic Torrents that contains submissions and comments from the top 40,000 subreddits in Reddit’s history (2005-06 to 2024-12).\added{To scale up upon the Study 1, we added an additional step to select more subreddits that well-suited for the broader research topic. As illustrated in Figure ~\ref{fig:collection-reddit}, we started with} collecting additional metadata for each subreddit through the Reddit API, including the number of members, primary language, the over18 label \added{(whether the subreddit contains adult-only content)}, and the public description section (a short description provided by moderators that explains the subreddit’s purpose, rules, or focus). 

We excluded subreddits that contains adult-only content and those whose primary language was not English, resulting in 25,691 subreddits. To avoid sampling from niche communities, we further restricted our analysis to the largest 20\% of subreddits by member count, resulting in a set of 5,138 subreddits.

\paragraph{Online Forum Selection}
We used an LLM (gpt-4o-mini-2025-03-01) to support the subreddit selection process. First, we randomly sampled 100 of the 5,138 subreddits and asked one researcher to rate their relevance on a 1–5 Likert scale, where 1 indicated not relevant and 5 indicated highly relevant. Relevance was assessed from two perspectives: (1) population relevance: whether members of the subreddit reasonably reflect the population of interest (in this case residents of the United States.); and (2) topic relevance: whether the discussions contain opinions related to the research topic. We did not expect any single subreddit to be a comprehensive representation of the targeting population, but we expected the combined set of selected subreddits to approximate this coverage. Next, we asked the LLM to label the same subreddits in the same format and to provide a rationale for each rating. The input included the subreddit’s name and the \dquote{description} section written by its moderators. We then compared the researcher’s labels and the LLM’s labels using Cohen’s Kappa, refining the prompts until the agreement reached 0.7 (the final prompt is attached in the Appendix~\ref{appendix:prompting}). In each iteration of this refinement process, we randomly sampled a different set of 100 subreddits for labeling to ensure that the evaluation did not overfit. As the dataset ultimately required a binary outcome (i.e., whether a subreddit was used as a data source), we converted the Likert ratings into 0/1 labels. The Likert scale at the labeling stage is to allow more flexibility for end users. After this adjustment, we applied the LLM to label all 5,138 subreddits. For the topic of this project, we consulted a public polling expert and set thresholds of 3 and 4 for the two relevance ratings. Specifically, subreddits with a member relevance rating of 3 or higher and a topic relevance rating of 4 or higher were included in the dataset. This results in a list of 288 relevant subreddits. We then filtered out data before 2023 to be more consistent with the interview data. 

\subsubsection{Data Source: Interview} 
\begin{figure}[htb]
    \centering
    \includegraphics[width=0.6\linewidth]{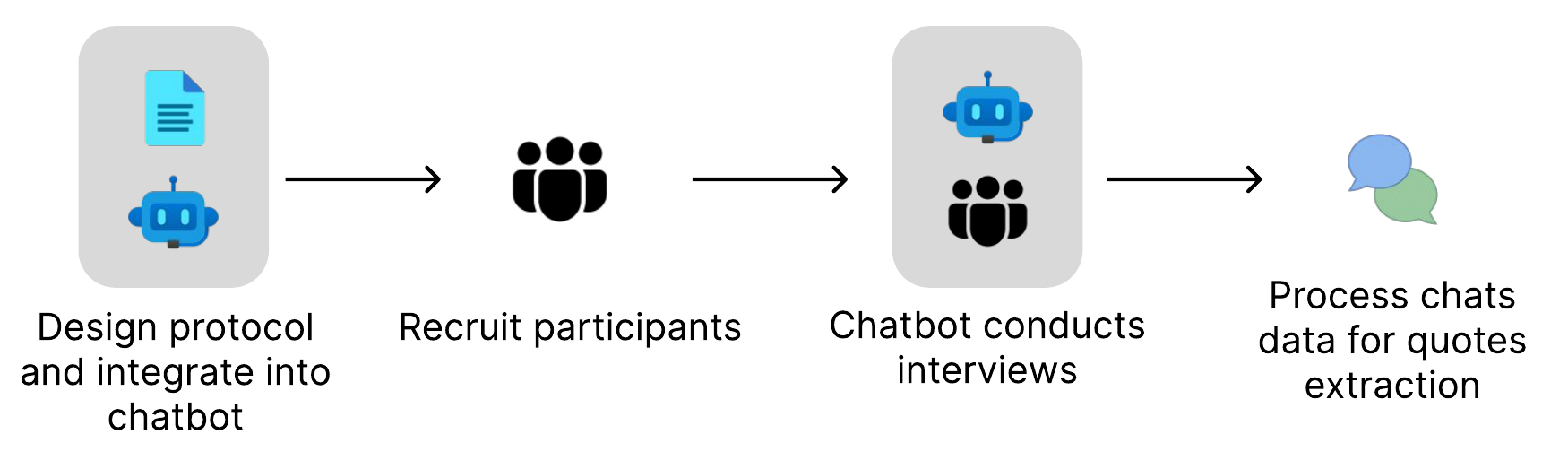}
    \caption{Workflow for collecting and processing chatbot-led interview data in Study 2. A public-opinion expert designs a semi-structured interview protocol focused on how people perceive the economic impacts of AI, and participants are recruited via Prolific. An LLM-based chatbot conducts 30-minute text interviews, after which participants complete a demographic questionnaire. We then process the interview transcripts using the same LLM-assisted quote-extraction and thematic analysis pipeline as for Reddit, enabling a direct comparison between themes derived from online forums, chatbot-led interviews, and authoritative policy reports.}
    \Description{Workflow for collecting and processing chatbot-led interview data in Study 2. A public-opinion expert designs a semi-structured interview protocol focused on how people perceive the economic impacts of AI, and participants are recruited via Prolific. An LLM-based chatbot conducts 30-minute text interviews, after which participants complete a demographic questionnaire. We then process the interview transcripts using the same LLM-assisted quote-extraction and thematic analysis pipeline as for Reddit, enabling a direct comparison between themes derived from online forums, chatbot-led interviews, and authoritative policy reports.}
    \label{fig:interview-workflow}
\end{figure}

\paragraph{Interview Protocols and Topics}
To be able to evaluate the thematic coverage of online forum data as employed in Study 1, in this study we additionally conducted online interviews with 1,058 participants and investigated how people in the US perceive the economic impact of AI. Figure ~\ref{fig:interview-workflow} illustrates the workflow of the data collection using interviews. We developed an LLM-based chatbot web application for conducting text-based interviews by adapting and modifying a previously validated online tool for qualitative interviews~\cite{geiecke2024conversations}. A public polling expert designed the interview around three research questions: 
\begin{enumerate}
    \item How do people perceive AI affects or will affect their personal economic well-being?
    \item What strategies do individuals use to cope with the economic changes caused by AI?
    \item What role do individuals think the government should play in responding to AI-driven economic changes?
\end{enumerate}

The research questions were later transformed into an interview protocol that instructed an LLM-based chatbot to led the conversation with human participants. See details of the prompt in Appendix \ref{appendix:chatbot_interview_protocol}

\paragraph{Web Application and Chatbot Development}
The application utilized Streamlit for the frontend presentation and the OpenAI API (gpt-4o-2024-05-13) for generating interview questions that followed a researcher-defined interview guide while dynamically responding to participants' answers.
The interviewer chatbot conducted semi-structured interviews based on the interview protocol. Each section included one specific opening question and a general description of relevant follow-up topics.
The chatbot was programmed to ask open, non-leading questions related to each interview section topic. Participants' responses were followed up with questions requesting additional details or examples to better understand their perspectives. Importantly, the chatbot was prompted to demonstrate cognitive empathy by acknowledging participants' thoughts and opinions without judgment and by avoiding the suggestion of possible answers. For example, we explicitly instructed the chatbot: \pquote{Do not suggest specific interventions... but you may ask a follow-up...}. This approach adhered to established guidelines for qualitative research interviews~\cite{dunwoodie2023qualitative, small2022qualitative}. After each participant's text input, the complete interview guidelines and chat history were provided to the LLM as a prompt to generate the chatbot's next utterance. 

\paragraph{Interview Participants}
A total of 1,058 adult U.S. residents participated in an online chatbot interview study, with a median age of 45 (age range 18 – 84, SD = 15.47). The participants (50 \% female, 48.3 \% male, 0.7 \% other, 1\% unknown) came from varied ethnic backgrounds (White: 62.5 \%, Black/African American: 13.7 \%, Hispanic/Latino: 9.1 \%, Asian: 6.2 \%, Native American: 1.3 \%, Unknown: 7.2\%) and different educational attainment levels (High school or less: 12.1 \%, Some college/Associate’s degree: 32.0\%, Bachelor’s degree: 34.4 \%, Graduate degree: 21.2 \%, Unknown: 0.3\%). An additional 73 participants entered the study but did not complete it (e.g., due to technical issues or voluntary withdrawal) and were therefore excluded from analysis. Participants were recruited through the online platform Prolific. 

\paragraph{Procedure and Ethical Statement}
The interviews are conducted in Aug 2025. Participants received a consent form detailing the study's purpose and data protection guidelines. The chatbot began interviews after receiving consent from participants. Each interview automatically ended after 30 minutes, or earlier if all sections had been covered and no new points were raised. Following the interview, participants completed a demographic questionnaire and several standardized personality measures. In total, the study session lasted approximately one hour, and participants who completed the study received 16 USD in compensation. 

Participants were explicitly instructed not to disclose private or sensitive information during the interview and to share only details about their lives that they would be comfortable posting in a public forum. Prior to analysis, all responses were screened to ensure participants’ responses did not contain personally identifiable information. All interview transcripts were anonymized by replacing participants’ names with only referring to individual participants via random numerical IDs and securely stored on a dedicated server. No personally identifiable information was retained in the dataset. The study was reviewed and approved by the Institutional Review Board of our organization (anonymized).

\paragraph{Interview Data Processing} To prepare the chat-based interview data for inputting into the workflow, we first segmented each conversation into question–answer (Q–A) pairs. Because the interviews followed a chatbot-question / human-answer structure, we used each chatbot message as the anchor for a new question unit because human participants may responded with multiple consecutive messages to answer a single chatbot question. In this case, we concatenated all consecutive human messages following a chatbot question until the next chatbot message appeared. This procedure ensured that each Q–A pair reflected a complete, semantically coherent human response to a single chatbot question. This processing step produced an initial dataset of 27,042 entries. On everage, each conversation consisted of 26.85 Q-A pairs (SD=10.21) and 1493.03 words (SD= 526.04).

\subsubsection{Quote Extraction}
We then input data from both sources into the workflow. For Reddit data, we input 5,491,991 data entries from 288 different subreddits and extracted 122, 191 relevant quotes. For interview data, we used an LLM to detect and remove transitional sentences(e.g., greetings) and messages that is not relevant to the study topic. After filtering, the final dataset contained 16,029 human messages (referred to as quotes for consistency). On average, each quote contained 22.1 words (SD = 19.83).
\subsubsection{Thematic Analysis}
Following the workflow explained in Section 3, we first identified high-level themes based on the research topic manually based on the discussions within the research team and classified quotes from both data sources accordingly. The resulting groups were: general economic outlook, AI’s impact on personal economic situation, personal strategies and adaptation to AI, opinions on government responsibility and action regarding AI. For online forum data, we introduced two additional labels: off-topic, to filter out irrelevant content and reduce noise in the output of the data collection pipeline; and others, to capture additional insights beyond the pre-defined high level themes. Note that we don't have this label for interview data as all conversations follows the interview protocol

\subsubsection{Report Generation}
After the LLM-based workflow created the output, two researchers manually reviewed the themes and further synthesized the results with the consultation of a public polling expert. Output themes and example quotes are included in Appendix~\ref{appendix:reports}. We did not conduct sentiment analysis and report corresponding trends for two reasons. First, our research questions focus on people’s perceptions of AI’s economic impact rather than on individual sentiment. Second, our analysis is based on theme extraction from sentence-level quotes. While some quotes may carry emotional overtones, they often represent partial expressions of sentiment and are insufficient to capture how individuals perceive broader trends.

\begin{table}[h]
\centering
\footnotesize
\begin{tabular}{lccccc}
\hline
  & Off-topic\footnotemark & \makecell{Personal Economic \\Situation\footnotemark} & 
   \makecell{Personal Strategies} & 
   \makecell{Government’s Role} & Others \\
\hline
Interview 
& 0 (0.00\%) 
& 6,557 (39.42\%) 
& 5,670 (34.09\%) 
& 4,402 (26.49\%) 
& 0 (0.00\%) \\
Reddit    
& 33,063 (27.06\%) 
& 44,283 (36.24\%) 
& 22,988 (18.82\%) 
& 1,640 (1.34\%) 
& 20,217 (16.55\%) \\
\hline
\end{tabular}

\caption{Distribution of each high-level theme category across interview and Reddit data}
\Description{The table shows the label distribution across interview and Reddit data. The columns are Off-topic, Personal Economic Situation, Personal Strategies, Government’s Role, and Others. For interview data: 10,413 off-topic, 6,557 personal economic situation, 5,670 personal strategies, 4,402 government’s role, and 0 others. For Reddit data: 33,063 off-topic, 44,283 personal economic situation, 22,988 personal strategies, 1,640 government’s role, and 20,217 others.}
\end{table}

\footnotetext{For Reddit data, \dquote{off-topic} refers to quotes that focus only on the economic situation or only on AI, but not both.}
\footnotetext{We shortened the high-level theme names for presentation; all labels are framed around the AI's impact}

\subsubsection{Comparison to Authoritative Reports}
To validate our workflow, we compared the themes in the generated reports with authoritative reports. In principle, we might have chosen a single benchmark report, but in practice we did not find any one document that was both comprehensive enough and closely aligned with our research questions about public views on AI and the economy. Instead, we assembled a complementary set of reports. We curated candidate reports through keyword searches for \dquote{economic impact of AI, public opinion} on Google and on the websites of well-known organizations such as academic institutions, professional polling organizations, and non-partisan research institutes. From this list, we excluded reports that reflected non-U.S. perspectives because our Reddit data extraction and interview recruitment both targeted U.S. participants(see Section 5.1.1 and 5.1.2 for details). Following this screening process, we identified six authoritative reports to serve as baselines:

\begin{enumerate}
\item A \href{https://www.cbsnews.com/news/poll-negativity-economy-job-market-artificial-intelligence/}{CBS
 News} report on public views of the economy, job market, and AI.
\item A \href{https://www.gallup.com/analytics/695033/american-ai-attitudes.aspx}{Gallup}
 report on American attitudes toward AI (focusing on the economic dimension).
\item An \href{https://www.apa.org/pubs/reports/work-in-america}{APA}
 report on workers' well-being in America.
\item A \href{https://www.pewresearch.org/internet/2023/04/20/ai-in-hiring-and-evaluating-workers-what-americans-think/}{Pew
 Research Center} report on AI in hiring and worker evaluation.
\item A \href{https://www.pewresearch.org/science/2025/09/17/how-americans-view-ai-and-its-impact-on-people-and-society/}{Pew
 Research Center} report on Americans’ views of AI and its societal impacts (filtered to the economy-related content).
\item A \href{https://reports.weforum.org/docs/WEF_Future_of_Jobs_Report_2025.pdf}{World
 Economic Forum} report on the future of jobs (focusing on U.S.-specific findings).
\end{enumerate}

Because none of the authoritative reports were based on the exact same topic of our study, two researchers read all six reports and manually extracted themes that were closely aligned with our research questions. The researchers iteratively discussed their interpretations until reaching consensus on a final set of 22 themes, as shown in Table ~\ref{tab:authoritative_themes} in the Appendix. These themes were then compared against the reports generated by the workflow. For each theme in the authoritative reports, researchers identified a similar or equivalent theme in the generated report when one existed.
Note that we did not evaluate the performance of the underlying LLM-assisted data analysis framework itself, as it is an established method that has already been validated in prior research~\cite{rao2024quallm}. Our focus in this work is solely on assessing its application and utility within the context of policy research.

\subsection{Findings}
\begin{figure}
    \centering
    \includegraphics[width=\linewidth]{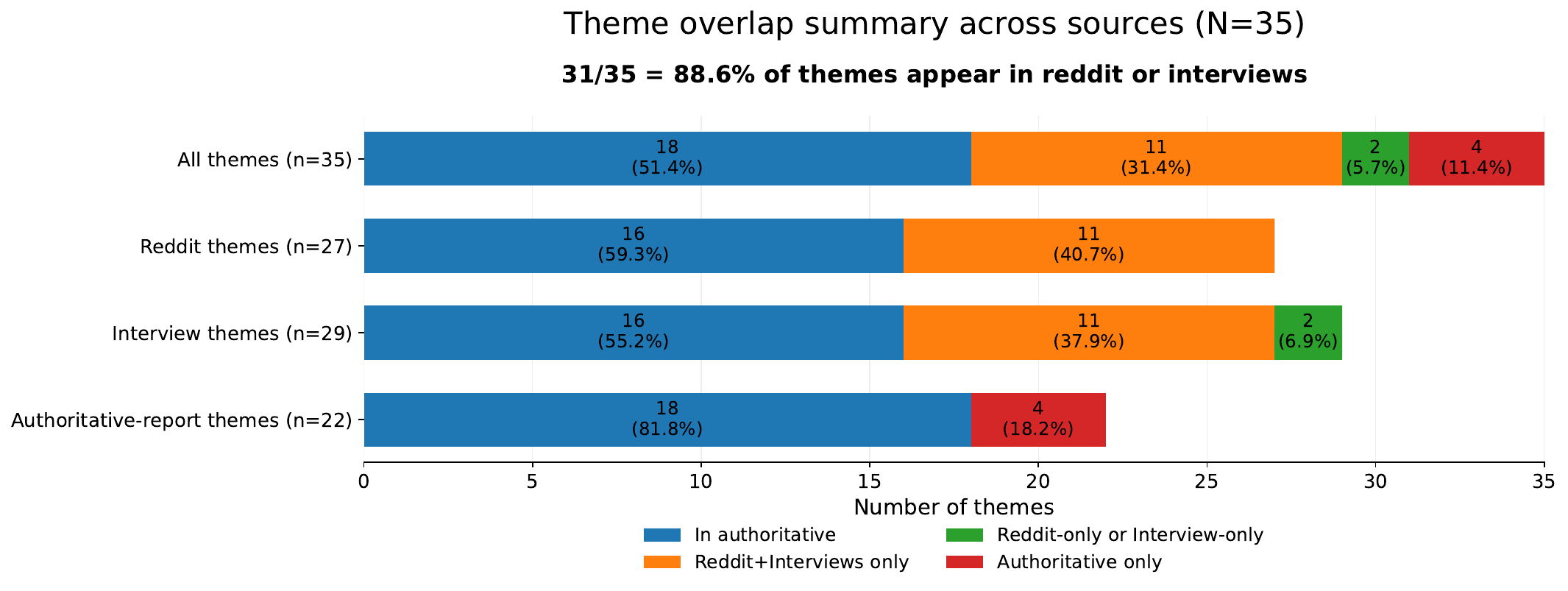}
    \caption{Theme overlap across sources. Reddit and interviews largely corroborate authoritative reports while adding additional themes. Across 35 themes, 31 (88.6\%) appear in Reddit or interviews, and most themes in authoritative reports also appear in these sources. The stacked bars show, for each source, how many themes are shared with authoritative reports versus how many are unique to Reddit/interviews or unique to authoritative reports. Refer to Table \ref{tab:theme_mapping} in the Appendix for a list of the themes.}
    \label{fig:overlap}
\end{figure}
Figure \ref{fig:overlap} provides an overview of how our themes align across sources. Table \ref{tab:framework-comparison} in Appendix lists the themes identified. Most themes surfaced in Reddit and interviews overlap with authoritative reports—suggesting that lived-experience data largely corroborates established concerns—while also introducing a smaller set of additional themes not captured in reports. In the remainder of this Findings section, we unpack these patterns in detail by comparing themes in (1) Reddit versus authoritative reports and (2) interviews versus authoritative reports.
\subsubsection{Themes from Reddit Data vs. Authoritative Reports}
As shown in Table ~\ref{tab:theme_mapping}, applying the workflow to Reddit discussions has high thematic coverage of themes in authoritative reports. Sixteen out of twenty-two themes identified in authoritative reports also appeared in the Reddit analysis, demonstrating that the workflow can reliably extract major public concerns from highly unstructured, informal text. Themes such as job-obsolescence concerns, expectations of major labor-market disruption, pessimism about the national economy, worries about inequality, and the perceived need for workforce retraining appeared consistently in both theme lists. Both authoritative reports and Reddit data also highlighted AI’s dual role as a productivity enhancer and a source of anxiety, as well as concerns about AI undermining creative thinking and calls for clearer disclosure of AI-generated content. This indicates that the workflow is capable of distilling coherent public issues that are typically surfaced through traditional policy research methods. 


Moreover, the workflow surfaced additional themes from Reddit data, including concerns rooted in community-specific anxieties. For example, the application of AI in freelancing and education is widely discussed in specific communities (i.e., subreddits) like r/uberdrivers and r/college, while in traditional surveys, such insights might be overlooked. In addition, the workflow detected some concrete policy suggestions, such as universal basic income and the protection of labor rights, which are not included in the authoritative reports. The emergence of these themes underscores the workflow’s ability to detect early signals and latent concerns that may not yet be incorporated into formal measurement instruments. Taken together, this combination of alignment and expansion highlights the potential of the workflow as a complementary and scalable analytic approach for policy research beyond traditional methods.

\subsubsection{Themes from Interview Data vs. Authoritative Reports}
By comparing the themes generated through our workflow with those identified in authoritative reports, we observed substantial overlap. The mapping results between themes in authoritative reports and interview data are shown in Table ~\ref{tab:theme_mapping}. Specifically, we found sixteen equivalent themes in the interview data out of the twenty-two themes extracted from authoritative sources, including concerns about job displacement, expectations of major labor-market disruption, pessimism about broader economic conditions, and the need for workforce retraining. Both sources also emphasized that AI could heighten inequality and place pressure on workers’ financial stability. In particular, themes produced by the workflow, such as mixed reactions toward workplace change, anticipated productivity gains from AI, are closely aligned with the themes emphasized in these reports. This strong alignment indicates that the workflow holds considerable promise for supporting policy research. Notably, we also identified differences. For example, several themes present in authoritative reports did not appear in the generated reports, such as concerns about mental health in the workplace and perceptions of fairness in AI-based hiring tools. Because these themes relate to specific domains that were not directly probed in our semi-structured interview protocol, we attribute their absence to the LLM-based chatbot not being sufficiently heuristic to surface cross-domain issues beyond the scope of the guided conversation. 

At the same time, the interviews revealed additional nuances, such as career rethinking strategies in response to the changes and the current uncertainty. These themes reveal the micro-level perspectives that complement the macro-level insights in authoritative reports from a personal perspective with specific examples. Our findings suggest that using the workflow for unstructured data like interview transcripts can serve as an efficient early-stage diagnostic tool that surfaces high-level public concerns typically captured by established methods.

\section{Discussion}

The findings in our work highlight the potential of LLM-assisted thematic analysis as a complementary method for policy researchers seeking rapid, cost-effective overviews of public opinions. At the same time, our analyses underscore the continued importance for ensuring and improving the thematic coverage of LLM-led workflows for policy research.

\subsection{LLM-based Tool Work as a Supplement to Traditional Policy Research}
Our study compared LLM-based workflow with authoritative reports produced with traditional policy research methods and evaluated its performance in extracting qualitative insights from unstructured texts. From the methodological perspective, traditional methods rely on carefully sampled populations, pre-defined surveys that are hard to scale up, often limited to several listening sessions and thousands of participants~\cite{britannica_opinion_poll, berinsky2017measuring, simmons1993capturing, hillygus2011evolution}. In our study, the workflow showed its potential to accommodate both small-scale qualitative interview corpora with thousands of entries and large-scale textual datasets with millions of entries and produced closely aligned results with authoritative reports using lower costs and shorter time. In addition, traditional policy research methods typically follow a deductive, top-down pipeline: experts define constructs, design question items, and interpret aggregate responses~\cite{rossi2013handbook}. As a result, the analytical space is bounded by what researchers decide to measure. In comparison, our LLM-assisted workflow operates in a more inductive, bottom-up manner. It codes themes at the level of individual quotes and then iteratively merges related concepts, mirroring qualitative thematic analysis and grounded theory practices~\cite{rao2024quallm}. In addition, traditional methods tend to use large-scale surveys that are designed for population-level generalizability, which pushes them toward broad, standardized questions that travel well across occupations and demographics, but leave less room for problems in specific communities that require special attention. The workflow, leveraging community-focused online forums(i.e. subreddits) is able to surface concerns that are highly contextual and community-specific. 

In the comparison, we found that methods used in producing the authoritative reports focus on quantifying public views towards a certain trend, while the themes identified by the workflow mainly capture the content of personal experience discourse rather than the emotional valence. Although unstructured data can support downstream quantitative tasks such as sentiment analysis, such results often suffer from ambiguity in natural language (e.g., sarcasm, double negatives, or emotionally mixed statements)~\cite{zhang2024performance, ortony2022all}. As a result, the workflow is especially well-suited for uncovering qualitative insights about what people are concerned about and how they reason through those concerns, rather than for producing sentiment proportions. This illustrates that LLM-based tools are not substitutes for traditional methods, but rather complementary analytic instruments that expand the range of research questions that can be addressed at both small and extremely large scales.

\subsection{Challenges of Trust and Interpretability in AI-Assisted Thematic Analysis}
Several policy researchers evaluating our framework voiced inherent distrust in AI analysis and requested methods to bridge the trust gap. This resistance to LLM-based methods appeared to stem from a general distrust of AI, not particularly our specific framework. As suggested by recent research, the lack of human intervention and transparency can exacerbate this distrust, as end users may feel uncertain about the process and its output~\cite{schneier2025ai, dwork2022distrust, laux2024institutionalised}. To mitigate these concerns, we incorporated human annotation and an explicit \dquote{approval} step during the data collection process, as well as manual review of a subset of extracted quotes to verify the legitimacy of the outputs. However, because the quotes were drawn from a masked dataset without original identifiers, we could not provide direct links back to source posts, which limited opportunities for participants to engage with the provenance of the data. Transparency might be improved by making data-processing steps explicit, clarifying how hallucinations are mitigated, and showing how extracted themes are validated~\cite {Afroogh2024TrustInAI}. User agency may be strengthened by allowing researchers to interactively review, filter, or revise LLM outputs rather than treating them as fixed summaries. Importantly, future interfaces should clearly communicate that AI-assisted thematic analysis is intended to complement, rather than replace, established qualitative methods. Collectively, these design strategies could address some of the concerns we identified in our study about credibility and control and help build a more sustainable and trustworthy workflow for AI-assisted qualitative summaries in policy research.

\subsection{Generalizing Beyond QuaLLM}
In this work, we adapt the QuaLLM framework into a policy research workflow using prompts tailored to policy research needs. Other LLM-based frameworks, such as TopicGPT and LLoom, could also be used at the thematic analysis stage, with adjustment like produce structured outputs and allow results to be traced back to the original text sources. This suggests that the workflow generalizes beyond a single framework and can accommodate evolving LLM-based analysis techniques. Although different LLM-based thematic analysis frameworks embed distinct assumptions about how themes are generated, they all follow the methodological principles of thematic analysis~\cite{braun2006using}, which have been shown to produce broadly aligned thematic outputs compared to traditional machine learning approaches or manual analyses, as demonstrated in prior evaluations~\cite{pham2024topicgpt, shankar2024docetlagenticqueryrewriting, lam2024concept}.

At the same time, as LLM-based workflows, particularly agentic designs, increasingly aim for high levels of automation, we emphasize that human oversight remains essential in policy research contexts. Our workflow explicitly identifies points where humans intervene, including defining analytic scope, selecting data sources, and interpreting prevalence rather than assuming end-to-end automation. This perspective suggests that the generalizability of LLM-based thematic analysis for policy research depends not only on model performance but also on how these systems are embedded within accountable and transparent research practices.

\section{Limitations and Future Work}
\label{sec:limitation}
This study has several limitations that should be acknowledged. First, the underlying data sources (Reddit posts in Study 1, Reddit posts and interviews in Study 2) might be constrained in their ability to surface public opinions from different subgroups of the population. Social media platforms such as Reddit may contain substantial biases, both in the sociodemographic characteristics of their users and in the types of topics that are discussed. Our interview data, although diverse with respect to age, gender, ethnicity, and educational attainment, also do not perfectly represent the United States population. Thematic outputs generated from these sources should therefore be interpreted as complementary qualitative insights rather than population-level estimates of opinion.

Second, our evaluation covered three exemplary policy-relevant topics, all within a United States context. The applicability and robustness of the framework may differ across policy domains, and cultural settings. Future work should evaluate LLM-assisted thematic analysis with additional topics and in international contexts, and examine how cultural and linguistic differences affect its usefulness for policy research.

Third, the qualitative interviews used in Study 2 were conducted via LLM-based chatbots rather than human interviewers. This allowed us to collect a substantially larger corpus than would have been feasible with human-led interviews, but the depth and nuance elicited by human interviewers may differ from those achieved by an LLM-based agent as discussed in \cite{cuevas2025collecting}. Future studies can directly compare thematic coverage derived from human-led and LLM-led interviews to clarify the trade-offs between scale and depth.

Fourth, although we employed several evaluation steps, the general reliability of LLM-generated summaries remains a central challenge, as reflected in the concerns voiced by expert users in our study. In this paper, the workflow required iterative prompt refinement until the agreement between researchers and the LLM reached 0.7, and this tuning process may introduce errors that propagate into subsequent labeling and extraction stages. Future research could better quantify and reduce such errors, potentially through enhanced model calibration or hybrid human–LLM validation mechanisms. In addition, future research can develop transparency-oriented interfaces and visualizations that help policy researchers critically assess the LLM identified themes. An important direction will be to explore how the current workflow can be improved and made more transparent, so that users not only understand how insights are produced but also feel confident in when and how to employ them for policy-relevant data analysis.

Finally, we focused on evaluating QuaLLM for policy research in this paper. As illustrated in Section 2.2, there are other LLM-assisted thematic analysis workflows. In this study, we demonstrate how we adapt the QuaLLM workflow. The adaptation has the potential to generalize to other LLM-assisted thematic analysis workflows by processing data from data source to fit the workflow, revising the prompts based on the data source and policy-relevant topic, and adjusting prompts according to the scale of input data. 

\section{Conclusion}
This study evaluated an LLM-assisted workflow in the context of policy research, examining its ability to support the analysis of public opinion across different policy-relevant topics and data sources. Through expert interviews and comparisons with authoritative human-authored reports, we found that the framework can recover many of the core themes identified by policy analysts while offering gains in speed and scalability. At the same time, our study highlights important limitations, including concerns about the representativeness of the underlying data and the reliability of LLM-generated summaries. These results suggest that LLM-assisted thematic analysis can serve as a useful complementary tool for policy researchers seeking rapid qualitative overviews, but it should not be viewed as a replacement for traditional methods of policy research. Human judgment and contextual expertise remain essential for deriving informed interpretations in a given policy area. Future work should further assess the conditions under which LLM-assisted approaches are most effective, benchmark them against alternative computational methods, and explore design strategies that improve transparency for users from non-technical backgrounds. By addressing these challenges, LLM-assisted frameworks may help integrate qualitative insights more efficiently into policy research and decision-making.

\section{Disclosure of the usage of LLM}
We used ChatGPT (GPT4o model\cite{openai2024gpt4o} and GPT5~\cite{openai2025gpt5}) to facilitate the writing of this manuscript. The usage includes:
\begin{itemize}
    \item Turn endpoints documentation into LaTeX format tables
    \item Correct grammar mistakes and spelling
    \item Polish the existing writing by prompts like "Find me a synonym of X", "What is the noun/adjective form of X" and "Shorten this sentence without changing its content".
\end{itemize}
\begin{acks}
    We thank our study participants for generously sharing their time and experiences. We also thank colleagues and collaborators for their helpful feedback and discussions throughout this project. This work was supported by the Nuremberg Institute for Market Decisions (Germany).
\end{acks}

\bibliographystyle{ACM-Reference-Format}
\bibliography{reference}


\appendix
\added{\section{Prompting \& Additional Technical Details}}
\label{appendix:prompting}
Built on a Flask backend with LLM-powered analysis, the workflow processes unstructured text to surface both high-level themes and supporting evidence. To optimize performance, the workflow caches processed reports for immediate future access and employs a modular architecture designed to scale across different data sources beyond its current focus on Reddit discussions. 

We used GPT-4 as our underlying LLM. The total cost to generate a complete report from 10,000 aggregated posts using the workflow is approximately $\$150-$\$300, varying based on the number of quotes aggregated and the length of the posts. For prompting, our primary aim was to minimize human intervention during the text analysis pipeline, resulting in prompts structured to generate high-level topic themes, categorize them into subtopics, and finally aggregate the results.

\subsection{Data Source Recommendation Prompt}
The current implementation of the Study 1's data source recommendation. As the workflow is specifically tailored to Reddit communities, serving as a proof-of-concept for our broader vision of public discourse analysis. We used the following prompt structure:

\begin{lstlisting}[basicstyle=\ttfamily\scriptsize]
Here is a list of subreddits: \{subreddits\_chunk\}. Based on the topic '\{topic\}', please provide a list of the most relevant 
subreddits from the list. If there are multiple relevant subreddits, separate their names with commas. If none are relevant, 
respond with a blank line.
\end{lstlisting}

While this implementation effectively serves our current focus on Reddit data, we designed the system's architecture with future extensibility in mind. The data source recommendation component is built as a modular system that can be expanded to incorporate diverse public data sources through the following architectural considerations

\subsection{Subreddit Labeling Prompt}
In Study 2, for identifying subreddits at scale to serve as a reliable data source, we designed a prompt that labels relevance about the population and content of each subreddit to filter the most approporiate ones:

\begin{lstlisting}[basicstyle=\ttfamily\scriptsize]
You are evaluating whether a subreddit should be included in a dataset for a public opinion study focused on \$target_group.
Please assess the subreddit based on the following two dimensions:
Who is posting: Do the users who post or comment in this subreddit represent a relevant population to include in a national public opinion sample, regardless of what they are saying? Focus only on who they are, considering indicators such as geographic location (e.g., U.S.-based), occupation, or other characteristics that make their perspectives relevant. The subreddit does not need to be demographically diverse; it can still be valuable as a subset of a broader dataset.
What is being said: Are the posts and comments in this subreddit valuable for understanding public opinion on $topic? Focus on whether the content reflects people's lived experiences, concerns, or reflections in ways that are accessible, grounded in everyday life, and potentially informative for policymakers. Both expert and non-expert perspectives may be useful.
Respond only with a valid JSON object. Do not include any explanations or formatting outside of the JSON. The output should have the following structure:
json
{{
    "population_relevance": integer from 1 to 5 (where 1 = not relevant at all, 5 = highly relevant),
    "population_reason": "Brief explanation of who they are and why they should or shouldn't be included",
    "content_relevance": integer from 1 to 5 (where 1 = not valuable at all, 5 = highly valuable),
    "content_reason": "Brief explanation of the kind of discussions and why they are or aren't useful for understanding opinions on $topic"
}}
The subreddit has the following metadata:
Subreddit name: \$subreddit
About description: \$about
\end{lstlisting}

\subsection{Theme Generation Prompt}
\label{appendix:theme generation prompt}
In Study 1, for generating high-level themes, we developed a prompt that emphasizes policy relevance and structural consistency:

\begin{lstlisting}[basicstyle=\ttfamily\scriptsize]
Generate a list of 9 themes that policy researchers would be interested in learning more about, related to the subreddit 
\{subreddit\}, each with a title \{'title'} and a very brief description \{'description'}. Return the themes in JSON format.
\end{lstlisting}

The prompt’s design focuses on targeted audience alignment, specifically mentioned policy researchers, structural requirements, JSON output format for consistent parsing, and scope control setting a specific number of themes to ensure comprehensive coverage without overwhelming users.

\subsection{Quote Analysis Prompt}
The quote analysis prompt represents are most complex prompt engineering effort, designed to extract meaningful policy insights dynamically based on the users’ selected data source and provided theme:

\begin{lstlisting}[basicstyle=\ttfamily\scriptsize]
You are analyzing data from the subreddit \{\$subreddit}, containing discussions about various aspects of \{\$topic}, including 
post titles, introductory text, and main content. The high-level theme we're interested in is \{\$theme}. Your task is to extract 
only the most relevant quotes, personal experiences, and opinions that explicitly mention or discuss concerns, risks, or 
implications related to \{\$theme\_focus}.

Please process each row and output only quotes that:
1. Directly reference \{\$theme\_focus} concerns.
2. Address specific risks, dangers, or ethical concerns related to \{\$concerns\_scope}.
3. Include personal anecdotes or experiences discussing \{\$theme\_focus} implications of \{\$topic}.

For each relevant quote, create an output entry in JSON format with the following structure:
{
  "entries": [
    {
      "quote": "Full quote of a personal experience or opinion explicitly mentioning $theme_focus concerns in $topic",
      "summary": "A brief summary of the quote, providing context or the main idea in less than 8 words"
    },
    {
      "quote": "Another relevant quote or anecdote about $theme_focus",
      "summary": "Summary or context of this second quote in less than 8 words"
    }
  ]
}
\end{lstlisting}

\begin{quote}
If no quotes or relevant content about \texttt{\$theme\_focus} concerns are found in the data, return \texttt{null}.
\end{quote}

This is a templated prompt designed to adjust for varying subreddit forums and theme focuses. The context setting defines the data source and scope of the analysis, establishes the analytical framework, and specifies the theme focus to guide the overall direction of the project. Moreover, the filtering criteria require direct references to concerns related to the chosen theme, prioritizing specific risks, ethical considerations, and personal anecdotes or experiences that provide deeper insight into the topic. Lastly, the output structure includes the extraction of quotes with full context, concise summaries limited to under eight words, and the preservation of source references to maintain credibility and traceability.
\subsection{Subtopic Identification Across Aggregated Quotes}
In Study 1, for analyzing aggregated quotes and identifying subtopics from the aggregated data of relevant quotes and summaries aligned with the defined theme, the workflow employs a prompt focused on subtopic analysis for further fine-grained insights:
\begin{lstlisting}[basicstyle=\ttfamily\scriptsize]
You are a research assistant helping to analyze summaries of \{\$subreddit} discussions.

Analyze the provided summaries and identify the top 9 most prevalent themes or codes.  
For each code:
1. Provide a clear, concise name. The theme name should be specific and not too broad.
2. Provide a brief description.

Respond in valid JSON format with the following structure:
{
    "codes": [
        {
            "name": "Theme Name",
            "description": "Brief description of what this theme represents."
        }
    ]
}
\end{lstlisting}
In Study 2, we adjusted the prompt to fit in large-scale data, in which we process data in batch (a batch contains 500 quotes). We then added an additional step to merge similar and duplicate themes with the prompt:

\begin{lstlisting}
[basicstyle=\ttfamily\scriptsize]
You are given multiple theme lists (JSON arrays), each derived from different batches of the same dataset.
Your job is to MERGE them into a single, deduplicated, rank-ordered list.

Output strictly as a JSON array of objects:
OUTPUT FORMAT (JSON ONLY)
{
    theme_title: string,
    theme_description: string of 10-20 words,
    sub_theme_ids: [index, ...]
}

Rules:
- Merge similar/overlapping themes into one broader theme.
- Keep themes sufficiently distinct. Merge close variants into one broader, canonical theme.
- Use ONLY the provided integer indices to build `sub_theme_ids`.
- Do NOT invent indices.
- EVERY input index MUST appear in EXACTLY ONE cluster. Do not miss any. Do not duplicate across clusters.
\end{lstlisting}

\subsection{Quote Categorization Prompt}
The final stage involves mapping quotes to subtopics using a prompt that emphasizes precise categorization:
\begin{lstlisting}[basicstyle=\ttfamily\scriptsize]
You are a research assistant helping to categorize quotes about \{\$subreddit} discussions. You will be provided with:
1. A numbered list of codes (1-9) with their descriptions.
2. A list of quotes to categorize.

For each quote, assign the ONE MOST appropriate code number (1-9) based on the themes present in the quote.

Respond in valid JSON format with the following structure:
{
    "categorized_quotes": [
        {
            "quote": "original quote text",
            "source_id": "original source id",
            "codes": [
                {
                    "code": code_number,
                    "code_name": "name of the assigned code"
                }
            ]
        }
    ]
}
Guidelines for categorization:
- Assign THE ONE MOST relevant code to each quote.
- Include the theme that is most substantively discussed in the quote.
- Be consistent in how you apply the codes.
- Use the code descriptions to guide your decisions.
- The one code in the \texttt{codes} array should represent a significant theme in the quote, not just minor mentions.
\end{lstlisting}
Note that in Study 2, we removed the constraint of 9 codes because the dataset is large.

\section{Interview Protocol for LLM-Based Chatbot}
\label{appendix:chatbot_interview_protocol}

In the interview, please explore the respondent’s personal economic outlook and their views on how artificial intelligence (AI) influences their economic situation and career. The interview consists of successive parts outlined below. Ask one question at a time and do not number your questions.

Begin the interview with:
‘Hello! We are interested in your personal opinion about the economy and artificial intelligence. Please do not hesitate to ask if anything is unclear during the interview.’

\textbf{Part I: General economic outlook}

Begin by exploring the respondent’s general view on the current state of the economy. Ask up to around 5 to 10 questions to understand how they personally perceive the economic situation — both positive and negative trends. Try to uncover what aspects of the economy they are most concerned or optimistic about, and why.
If the respondent starts talking about specific technologies or artificial intelligence here, acknowledge their point but gently guide the conversation back to their general economic views in this part.
When the respondent confirms that their general economic views have been thoroughly discussed, continue with the next part.

\textbf{Part II: Impact of AI on personal economic situation}

Next, explore whether and how the respondent believes that artificial intelligence will influence their personal economic well-being — for example, their job, career, or financial situation.

Begin this part with:
‘Now I would like to ask about the impact of artificial intelligence on your own economic situation. Do you feel like AI is already affecting you, or might do so in the future?’

Ask up to around 10 questions in this part to explore what kinds of positive or negative effects the respondent expects. Encourage concrete examples and reasons. Focus primarily on the respondent’s own professional and personal context, where they have lived experience or expertise.
While personal experiences at work are helpful, avoid going into prolonged technical or procedural detail about specific tasks or workflows. Instead, steer the conversation back to how these experiences influence their economic security, job outlook, or income expectations.
If the respondent raises broader topics (e.g. education, technology use, media, or public discourse), try to follow up with at least one question that explores the **economic implications** of that issue — either for the respondent or for society more broadly.

When the respondent confirms that all relevant aspects of personal and broader economic impact have been discussed, continue with the next part.

\textbf{Part III: Personal strategies and adaptation}

Explore how the respondent personally plans to cope with or prepare for economic changes related to AI.
Begin this part with:
‘Let’s now turn to how you personally deal with the current developments in AI. Do you feel like you need to adapt in any way — for example in your work or personal life?’
Ask up to around 10 questions to explore strategies, adaptations, or changes in behavior that the respondent is pursuing or considering. Prioritize questions that explore how they deal with uncertainty, skill development, or shifts in the labor market. Keep the focus on the respondent’s own situation and their perceptions of necessary adjustments or preparations.

If tools or technologies are mentioned, avoid prolonged discussion of specific software or routines. Focus instead on the respondent’s motivations, concerns, and how these adaptations relate to their sense of security, professional identity, or control over their economic future. If the respondent expresses that they do not feel like they need to adapt in any way, consider asking how they think that approach might affect their economic position or opportunities in the long term.

When the respondent confirms that their personal strategies have been fully discussed, continue with the next part.

\textbf{Part IV: Government responsibility and action}

Explore what the respondent believes the government should or should not do to address the challenges and opportunities associated with AI.

Begin this part with:
'Finally, I’d like to ask about what role you think the government should play in preparing society for AI. Do you think any political action is needed — and if so, what kind?'

Ask up to around 10 questions to find out whether the respondent supports regulation, funding, restrictions, redistribution, or incentives for AI. Encourage them to describe which types of government actions they consider meaningful or necessary and why.

Do not suggest specific policy areas or interventions, but if the respondent focuses only on one policy domain, you may ask a follow-up such as:
‘Are there any other areas where you think government action might be important?’

When the respondent raises broader concerns (e.g. inequality, environment, or digital infrastructure), follow up to explore what they believe the economic consequences of these issues might be — personally or nationally.

Once the respondent confirms that all aspects of their views on political action have been thoroughly discussed, you may end the interview or return to any earlier topic if needed.

\added{\section{Study 1: Evaluation Script}}
\label{appendix:evaluation}
\subsection{Pre-Task Questions} 
\begin{enumerate}
    \item How many years of experience do you have in policy research or analysis?
    \begin{itemize}
        \item In your current role, how often do you need to gather public opinion data? (Daily, Weekly, Monthly, Rarely, Never)
        \item How confident are you in your ability to quickly gather public opinions on a topic using online tools? (Scale 1-5, where 1 = Not confident at all, 5 = Extremely confident)
    \end{itemize}
    \item Topic Familiarity "Please rate your familiarity with the following topics:" (Scale 1-5, where 1 = Not familiar at all, 5 = Very familiar). How do you typically stay informed about these topics?
    \begin{itemize}
        \item Social Media Impact On Children
        \item Climate Change
    \end{itemize}
    \item AI Tool Experience
    \begin{itemize}
        \item Have you used any AI-assisted research tools before? 
        \begin{itemize}
            \item If yes, which ones?
            \item How frequently do you use AI-assisted research tools?
        \end{itemize}
        \item What are your expectations for how AI might help in gathering public opinion?
    \end{itemize}
    \item For your most recent policy research project:
    \begin{itemize}
        \item Roughly how many weeks did it take from start to finish?
        \item What percentage of that time was spent specifically on gathering public sentiment?
        \item Please sketch out your typical workflow when researching public opinion on a policy topic. Include all steps from initial research to final insights
        \begin{itemize}
            \item Estimate time spent on each step
            \item Mark which steps you find most time-consuming or complex
        \end{itemize}
    \end{itemize}
\end{enumerate}

\subsection{Task Procedure}
Phase 1 (10 minutes): "You'll be assigned to either Topic A (Social Media and children) or Topic B (Climate Change). Your goal is to gather as many meaningful insights about public opinion on this topic as possible in 10 minutes. Try to identify both specific anecdotes and broader themes in public discourse.

[For AI-assisted group]:
\begin{itemize}
    \item You'll be using an AI research tool, which can help analyze social media posts and online discussions
    \item The tool will show you relevant posts and help identify common themes
    \item You can use the tool's features to filter and categorize the information
    \item Please take notes on any insights you find using the given template
\end{itemize}

[For traditional method group]:
\begin{itemize}
    \item You'll have access to standard web browsers and search tools
    \item You can visit any public websites, forums, or social media platforms
    \item You may not use any AI tools such as ChatGPT
    \item Please take notes on any insights you find using the given template
\end{itemize}
Remember to focus on both individual stories and broader patterns in public opinion."
Phase 2 (10 minutes): "Now you'll switch to the other topic and use the other research method. The same goals apply - gather as many meaningful insights as possible in 10 minutes."

\subsection{Post-Task Survey}
"Now that you've completed both tasks, please share your experience with each method."
Quantitative Metrics:
\begin{enumerate}
    \item For each topic and method, please indicate in the given template:
    \begin{itemize}
        \item Number of distinct anecdotes gathered
        \item Number of broader themes identified
        \item Time spent finding your first meaningful insight
        \item Confidence in the comprehensiveness of your findings (Scale 1-5
    \end{itemize}
\end{enumerate}

\textbf{Process Comparison}
“Now that you have used the AI tool, sketch your research workflow using the LLM-workflow for the same research task." 
\begin{itemize}
    \item Write the new process  (real quotes and testimonies) 
    \item Mark steps that were:
    \begin{itemize}
        \item Eliminated completely
        \item Simplified/compressed
        \item Unchanged
    \end{itemize}
    \item For simplified steps: Briefly explain how LLM-workflow changed the step
\end{itemize}

\textbf{Comparative Questions (5-point scale)}
Rate how much faster or slower LLM-workflow was for:
\begin{itemize}
    \item Finding relevant public opinions (Much slower 1-2-3-4-5 Much faster)
    \item Understanding sentiment trends (Much slower 1-2-3-4-5 Much faster)
    \item Generating summary reports (Much slower 1-2-3-4-5 Much faster)
\end{itemize}
Which method was more effective for:
\begin{itemize}
    \item Finding specific examples/anecdotes?
    \item Identifying broader themes?
    \item Understanding the range of public opinions?
    \item Discovering unexpected insights? (Rate each on a scale 1-5 for both methods
\end{itemize}

User Experience: For the AI-assisted tool:
\begin{itemize}
    \item How intuitive was the interface? (Scale 1-5)
    \item What features were most helpful?
    \item What features were missing or could be improved?
    \item Did you trust the tool's analysis? Why or why not?
\end{itemize}

For the traditional method:
\begin{itemize}
    \item What strategies did you use to find information quickly?
    \item What were the main challenges you encountered?
    \item What advantages did this method have over the AI tool?
\end{itemize}

Comparative Analysis: Please compare the two methods:
\begin{itemize}
    \item Which method would you prefer for future research tasks? Why?
    \item How did the effectiveness of each method vary between the two topics?
    \item What would be your ideal combination of AI-assisted and traditional research methods?
\end{itemize}

\subsection{Interview Worksheet} \label{sec:Interview Worksheet}
Topic: [Social Media’s Impact on Youth/Climate Change]. Method: [Traditional or AI-Assisted]

\textbf{Themes Identified}
\begin{enumerate}
    \item Theme: 
    \begin{itemize}
        \item Brief Description:
        \begin{itemize}
            \item Describe the main stakeholder perspectives
            \item Summarize proposed solutions or interventions mentioned in the discussions
            \item Note any emerging trends
            \item Note significant disagreement or consensus
        \end{itemize}
        \item Supporting Evidence (2-3 examples): 
        \begin{itemize}
            \item Can be an anecdote/story
        \end{itemize}
    \end{itemize}
        \item Theme: 
    \begin{itemize}
        \item Brief Description:
        \item Supporting Evidence (2-3 examples): 
    \end{itemize}
\end{enumerate}
[Continue for each theme...]

\textbf{Quick Statistics}
\begin{itemize}
    \item Total Themes Identified: [Count]
    \item Total Anecdotes Collected: [Count]
    \item Time to First Insight: [Minutes]
\end{itemize}

\textbf{Notes}
[Any quick observations about the research process or findings]

\clearpage
\section{Study 2: Themes Identified in Authoritative Reports and Descriptions }

\begin{table}[hbt]
\centering
\footnotesize
\begin{tabular}{p{7.7cm} c c c}
\hline
\textbf{Themes} & \textbf{Authoritative Reports} & \textbf{Reddit} & \textbf{Interview} \\
\hline
Mixed Sentiments Toward the Changes & Y & N & Y \\
Positive Impact of Working in Preferred Environment & Y & Y & N \\
Value Psychological Well-Being at Work & Y & Y & N \\
AI Adoption Leads to Higher Productivity & Y & Y & Y \\
Job-Obsolescence Concerns & Y & Y & Y \\
Persistent Pessimism About the National Economy & Y & Y & Y \\
Personal Financial Stability Amid Growing Inequality & Y & Y & Y \\
Perception of a Weak Job Market With Fewer Quality Jobs & Y & Y & Y \\
Job Security Paired With Pessimism About Mobility & Y & Y & Y \\
AI Seen as a Net Threat to Jobs in One’s Field & Y & Y & Y \\
Desire for Support in Workforce Retraining & Y & Y & Y \\
More Funding for AI Research & Y & N & N \\
AI-Related Tax Incentives & Y & N & N \\
Reducing Regulatory Barriers to AI Development & Y & Y & Y \\
Expectation of Major Labor-Market Disruption & Y & Y & Y \\
Lower Perceived Personal Impact of AI & Y & N & N \\
Opposition to AI in High-Stakes Workplace Decisions & Y & N & Y \\
AI Viewed as Potentially Reducing Bias in Hiring & Y & N & N \\
Concerns About AI Undermining Creative Thinking & Y & Y & Y \\
Support for Clear Disclosure of AI-Generated Content & Y & Y & Y \\
Perception of Rapid Growth in AI-Related Jobs & Y & Y & Y \\
Active Efforts to Learn AI-Related Skills & Y & Y & Y \\
Salary and Compensation Issues & N & Y & Y \\
Freelancing and Gig Economy Challenges & N & Y & Y \\
Financial Planning and Stability & N & Y & Y \\
Universal Basic Income (UBI) & N & Y & Y \\
Support for Labor Rights & N & Y & Y \\
Economic Policy and Reform & N & Y & Y \\
AI in Specific Industries & N & Y & Y \\
AI Hype and Skepticism & N & Y & Y \\
Outsourcing and Global Competition & N & Y & Y \\
Investment Opportunities in AI & N & Y & Y \\
Career Transition and Adaptation & N & Y & Y \\
Coping with Uncertainty & N & N & Y \\
Skepticism Towards Government Involvement & N & N & Y \\
\hline
\end{tabular}
\caption{Comparison between Themes in Authoritative Reports vs. Reddit and Interview Data. Y means this theme appears in the report and N means it does not.}
\label{tab:theme_mapping}
\end{table}

\begin{longtable}{p{1cm} p{4.3cm} p{7.5cm}}
\hline
\textbf{Source} & \textbf{Theme} & \textbf{Description} \\
\hline
\endfirsthead

\hline
\textbf{Source} & \textbf{Theme} & \textbf{Description} \\
\hline
\endhead
2,3 & Mixed Sentiments Toward Organizational Change & People express both positive and negative emotions toward workplace changes, with stress and anxiety more common among people who are dissatisfied with their compensation. \\

3 & Positive Impact of Working in Preferred Environment & People who work in their preferred arrangement report better mental health, higher job satisfaction, and a stronger sense of belonging. \\

3 & Value Psychological Well-Being at Work & People place high value on organizations supporting their emotional and psychological well-being, though satisfaction with available mental-health resources varies. \\

1, 3, 5 & AI Adoption Leads to Higher Productivity & People who feel more productive than last year often attribute these gains to AI or other technologies and report stronger morale and healthier workplace relationships. \\

3 & Job-Obsolescence Concerns & People are concerned that AI may make some of their job tasks unnecessary. \\

1 & Persistent Pessimism About the National Economy & People believe the current economy is performing poorly and expect conditions to worsen. \\

1 & Personal Financial Stability Amid Growing Inequality & People feel personally financially stable yet increasingly strained, with widening gaps between higher- and lower-income households. \\

1 & Perception of a Weak Job Market With Fewer Quality Jobs & People believe the job market now offers fewer high-quality positions than it did five years ago. \\

1 & Job Security Paired With Pessimism About Mobility & People feel secure in their current job but doubt they could easily find a similar or better job if they needed to search. \\

1, 2, 4, 6 & AI Seen as a Net Threat to Jobs in One’s Field & People expect AI to reduce job opportunities within their occupation over the next decade. \\

2 & Desire for Support in Workforce Retraining & People want more investment in programs that help them reskill or transition to different fields in response to AI-related changes. \\

2 & More Funding for AI Research & People support increased funding for AI research to strengthen technological leadership. \\

2 & AI-Related Tax Incentives & People favor tax incentives that encourage AI innovation and adoption. \\

2 & Reducing Regulatory Barriers to AI Development & People support simplifying regulations to accelerate responsible AI development. \\

4 & Expectation of Major Labor-Market Disruption & People believe AI will significantly reshape job roles, hiring practices, and workplace structures over the next 20 years. \\

4 & Limited Perceived Personal Vulnerability to AI Changes & People believe AI will have a broad societal impact but perceive its effect on their own jobs to be minor. \\

4 & Opposition to AI in High-Stakes Workplace Decisions & People oppose the use of AI for firing, promotions, performance evaluations, or workplace surveillance. \\

4 & AI Viewed as Potentially Reducing Bias in Hiring & People believe AI may improve fairness in hiring and evaluations in contexts where racial or ethnic bias is a concern. \\

5 & Concerns About AI Undermining Creative Thinking & People believe AI will weaken, rather than strengthen, their ability to think creatively. \\

5 & Support for Clear Disclosure of AI-Generated Content & People want transparent disclosure whenever content is generated or substantially assisted by AI. \\

6 & Perception of Rapid Growth in AI-Related Jobs & People believe jobs connected to AI and automation are expanding quickly and represent a major area of future employment growth. \\

6 & Active Efforts to Learn AI-Related Skills & People are actively acquiring AI-related skills to stay competitive and adapt to changes brought by technological transformation. \\

\hline
\caption{A List of Themes from Authoritative Reports about Public Perception on the Economic Impact of AI.}
\label{tab:authoritative_themes}
\end{longtable}
The Source column in the table refers to the ID of reports we identified the theme from. The reports are  
\begin{enumerate}
\item \added{A \href{https://www.cbsnews.com/news/poll-negativity-economy-job-market-artificial-intelligence/}{CBS
 News} report on public views of the economy, job market, and AI.}
\item \added{A \href{https://www.gallup.com/analytics/695033/american-ai-attitudes.aspx}{Gallup}
 report on American attitudes toward AI (focusing on the economic dimension).}
\item \added{An \href{https://www.apa.org/pubs/reports/work-in-america}{APA}
 report on workers' well-being in America.}
\item \added{A \href{https://www.pewresearch.org/internet/2023/04/20/ai-in-hiring-and-evaluating-workers-what-americans-think/}{Pew
 Research Center} report on AI in hiring and worker evaluation.}
\item \added{A \href{https://www.pewresearch.org/science/2025/09/17/how-americans-view-ai-and-its-impact-on-people-and-society/}{Pew
 Research Center} report on Americans’ views of AI and its societal impacts (filtered to the economy-related content).}
\item \added{A \href{https://reports.weforum.org/docs/WEF_Future_of_Jobs_Report_2025.pdf}{World
 Economic Forum} report on the future of jobs (focusing on U.S.-specific findings).}
\end{enumerate}
\section{Study 2: Reports Generated by the Workflow}
\label{appendix:reports}
\begin{longtable}{p{2.4cm} p{3.8cm} p{5.4cm} p{0.8cm}}
\hline
\textbf{Theme} & \textbf{Description} & \textbf{Quote} & \textbf{Ratio} \\
\hline
\endfirsthead

\hline
\textbf{Theme} & \textbf{Description} & \textbf{Quote} & \textbf{Ratio} \\
\hline
\endhead

\multicolumn{4}{l}{\textbf{ AI’s impact on personal economic situation(44,283)}} \\ \hline

AI Competition and Employment
& Discussions on AI's impact on job security, competition, and the future of various professions.
& ``The trend will be that fewer, good SWEs who can leverage these tools will be retained and the entry point will get harder.''
& 34.60\% \\ \cline{1-4}

Personal Financial Pressures
& Discussions on economic challenges, job transitions, and the impact of inflation on financial stability.
& ``AI already took my job. The AI Google results update tanked my blog’s traffic 90\%, which was my primary source of income. I’ve since started an in-person cleaning service. It should be a while before AI replaces me now.''
& 15.10\% \\ \cline{1-4}

Emotional Impact of AI-driven Changes
& Concerns about job security, layoffs, and the emotional toll of economic instability.
& ``Couple with AI plus general downturns from the streaming model ... it's just so stressful. I'm in a pretty good place personally, but the existential angst feels every present regardless. I really feel for those who are really suffering at the moment.''
& 9.70\% \\ \cline{1-4}

Salary and Compensation Issues
& Concerns regarding salary expectations, disparities, negotiations, and economic pressures on compensation.
& ``There are two groups of people who will be impacted the most: Entry-level data scientists and FAANG data scientists -- because their salary expectations are likely going to have to change dramatically.''
& 8.10\% \\ \cline{1-4}

Future of Work and Job Market
& Speculations on how AI will reshape job roles, market dynamics, and future career viability.
& ``If all jobs are going to get replaced then what will the human population do? ... AI is going to take a significant amount of jobs in all industries and it's very unlikely there will be enough jobs to go around.''
& 6.70\% \\ \cline{1-4}

AI in Creative Fields
& Comments about AI's impact on creative professions and job quality.
& ``Artists will only be kept until they've fed the machine enough art that the machine can do the job without them.''
& 4.90\% \\ \cline{1-4}

Work-Life Balance and Job Satisfaction
& Discussions on the importance of work-life balance, job satisfaction, and workplace conditions.
& ``Between AI, my own ineptitude and malcontentment, and the piling bills, I feel totally drained of hope for the career I wanted or even one that will make life feel bearable.''
& 4.20\% \\ \cline{1-4}

Freelancing and Gig Economy Challenges
& Insights on the struggles and opportunities faced by freelancers and gig workers in the current market.
& ``I do data annotation. It's boring as hell but pays decently for what it is. I believe there's a dedicated subreddit as well. A lot of the tasks consist of evaluating the responses from two AI language models and providing a short report on which one is better and why. It pays about \$20/hr USD to your PayPal.''
& 3.70\% \\ \cline{1-4}

AI Tools and Career Advancement
& Exploration of how AI tools can enhance productivity, job applications, and career growth.
& ``I saw a buddy of mine using AI to tailor his resume to the one job he really cared about, and it actually landed him an interview in this job market.''
& 1.70\% \\

\hline
\multicolumn{4}{l}{\textbf{ Personal Strategies to Cope with AI's Economic Impact(22,988)}} \\ \hline

Career Transition and Adaptation
& Strategies for navigating job changes, adapting to new roles, and coping with market shifts.
& ``Given the rapid advancements in AI and its impact across industries, I'm seriously considering transitioning to become an ML engineer.''
& 42.50\% \\ \cline{1-4}

AI Integration and Productivity
& Exploration of how AI tools are integrated into work processes to enhance productivity and efficiency.
& ``I'm automating my current businesses with AI. Then, once this becomes passive income, I'll flow the money into this bigger project, sell the house, and enjoy keeping 100\%.''
& 20.80\% \\ \cline{1-4}

Adapting Skills for AI
& Focus on learning and adapting skills to meet the demands of AI advancements in the job market.
& ``My ultimate goal is to get a deep understanding of AI, ideally with enough skills to potentially start something innovative in the field one day.''
& 15.50\% \\ \cline{1-4}

Financial Planning and Stability
& Strategies for managing finances, achieving stability, and preparing for economic uncertainties.
& ``Nowadays, job uncertainties are increasing. It's wise to save money and invest in something that can make us self-sufficient. By this, I mean we should be able to sustain ourselves indefinitely, even if there are no jobs available in the future.''
& 9.10\% \\ \cline{1-4}

Education and Skill Development
& Focus on the importance of education, certifications, and continuous learning for career advancement.
& ``I am considering pursuing a STEM PhD to have easier access to head of AI kind of jobs which today get literally saturated with PhD applicants.''
& 5.10\% \\

\hline
\multicolumn{4}{l}{\textbf{ People's Perceptions of Government’s Role in AI-Driven Economic Change(1,640)}} \\ \hline

Universal Basic Income (UBI)
& Discussion on the necessity of UBI as a response to job displacement caused by AI.
& ``One day all jobs will be automated. There will be two options, either a universal basic income based socialist/star trek like society, or a complete collapse of the economy making all the `wealth' of the ultra rich worthless and the world becoming basically Mad Max.''
& 28.40\% \\ \cline{1-4}

AI Regulation and Oversight
& Discussion on the need for government regulation to manage AI's impact on jobs and society.
& ``I believe regulation needs to be utilized, because AI could become a trillion dollar industry depending on how rapidly they advance, and we need to create safeguards to ensure that AI does not cause problems in the future that we won't know how to handle.''
& 20.10\% \\ \cline{1-4}

Support for Labor Rights
& Advocacy for labor unions and stronger protections for workers in the context of AI advancements.
& ``I am usually in favor of leaving the market alone but the Sherman Act definitely needs to make a swing through… Americans slowly waking up to the reason our grandparents were in unions… It's an inevitable economic problem; the destruction of unions in a capitalist system, combined with the rise of automation, will only serve to worsen conditions for the working class.''
& 7\% \\ \cline{1-4}

Economic Policy and Reform
& Discussion on the need for economic policies and reforms in response to AI and automation.
& ``We may need FDR like policies not because of inflation but due to what AI will do to the economy and job market. There will be jobs that will go the way of the type writer repairman. A lot of jobs are going to become obsolete and low and high skilled workers are going to suffer.''
& 6.40\% \\ \cline{1-4}

Education and Retraining for AI
& Emphasis on the importance of education and retraining programs to adapt to changes brought by AI.
& ``Taking jobs away and giving people money is not going to be enough. There needs to be an entirely new education system that tackles that issue.''
& 1.60\% \\

\hline
\multicolumn{4}{l}{\textbf{ Others(20,217)}} \\ \hline

AI Tools and Automation
& Exploration of AI tools and their role in automating processes and enhancing productivity.
& ``I'm curious how AI tools might streamline this process. What are your thoughts on using AI-powered knowledge management systems to help organize and optimize all this content creation? Could it potentially improve productivity or free up time from constantly being online?''
& 16.40\% \\ \cline{1-4}

AI Content Quality and Ethics
& Discussions on the quality, effectiveness, and ethical implications of AI-generated content.
& ``But this guy's point about AI being used by malicious actors to blur the lines [further] between reality and propagandic fantasy did make me stop and think for a bit.''
& 13.70\% \\ \cline{1-4}

AI in Education
& Exploration of AI's role in transforming educational practices, learning tools, and academic methodologies.
& ``Yeah there is going to be no incentive for a lot of people to learn a lot of things. Which on some level is okay if it's not run completely by for profit companies or government.''
& 9.40\% \\ \cline{1-4}

AI Hype and Skepticism
& Debates on the sustainability of AI trends and their real-world impact on society.
& ``AI \& ML is a road to broken dreams, the hype exists to milk dumb investment funds don’t get pulled down yourself. FAANG had stupid levels of hype, eventually their salaries \& promises came crashing down too. Hype surrounds unsustainable ideas.''
& 5.80\% \\ \cline{1-4}

Outsourcing and Global Competition
& Concerns about outsourcing jobs, its financial effects, and competition from international workers.
& ``In the future I doubt it'll be outsourced to other countries it'll probably outsourced AI when they get that perfected. It's all about the profitability and the cheapest labor that they can get.''
& 5.40\% \\ \cline{1-4}

Economic Pressures and Challenges
& Concerns about wage stagnation, rising costs, and economic pressures affecting various industries.
& ``I'm starting to think we might be forced into a revolution - if enough people can't afford rent and go broke with their backs up against the wall - well it won't be good. The cost of living is too high and you can only push so many people over the edge until it's too many and they start fighting back.''
& 4.50\% \\ \cline{1-4}

AI Accessibility and Affordability
& Concerns regarding the cost and access to AI technologies, impacting various users and industries.
& ``In the medium-term, only people with access to the compute power will be empowered by AI. So many people are going to be left in the dust like farmers in Eritrea, migrants in Nicaragua, etc. etc. It will also increase the power gap between oppressors and oppressed groups like the Rohingya or Uyghur.''
& 3.90\% \\ \cline{1-4}

Investment Opportunities in AI
& Exploration of investment strategies and opportunities related to AI technologies and sectors.
& ``Bank of America put out a piece saying that they expect software companies to increase AI investment significantly. They said that currently, software companies are spending around 4\% of their revenue on AI investment.''
& 2.10\% \\ \cline{1-4}

AI Integration and Implementation Challenges
& Discussions on the difficulties and costs associated with integrating AI into products and services.
& ``I have a product where we are working to integrate AI into it and it performs great 95\% of the time. But that 5\% is BAD, really bad, to the point where we had to scale back use cases because the risk wasn’t worth it.''
& 1.80\% \\
\hline
\caption{Generated Report: Reddit Data}
\end{longtable}

\begin{longtable}{p{2.4cm} p{3.8cm} p{5.4cm} p{0.8cm}}
\hline
\textbf{Theme} & \textbf{Description} & \textbf{Quote} & \textbf{Ratio} \\
\hline
\endfirsthead

\hline
\textbf{Theme} & \textbf{Description} & \textbf{Quote} & \textbf{Ratio} \\
\hline
\endhead

\multicolumn{4}{l}{\textbf{ AI’s impact on personal economic situation (6,557)}} \\ \hline

Job Displacement Concerns
& Participants express fears about AI replacing jobs across various sectors, leading to economic instability.
& ``A lot of companies are forgoing the cost and inconvenience of hiring and training new workers in favor of AI. that affects the job market for those like me.''
& 34.90\% \\ \cline{1-4}

AI as a Productivity Tool
& Participants see AI as a tool that enhances productivity and efficiency in their work across various sectors.
& ``Yea, AI has impacted my job and career in a positive way. Tasks that naturally take me like 4 hours to complete, with the help of AI, I can conveniently complete in 2 hours or less.''
& 15.20\% \\ \cline{1-4}

AI in Specific Industries
& Discussions on how AI is impacting specific industries like healthcare and education.
& ``Lesson planning can be done by AI now.''
& 14.90\% \\ \cline{1-4}

Broader Economic Impact of AI
& Participants discuss broader economic implications of AI, including money flow and inequality.
& ``So many roles in the labor market will be rendered obsolete, leaving less money flowing through the economy.''
& 7\% \\ \cline{1-4}

Future Job Market Predictions
& Participants speculate on how AI will shape future job markets, including potential new roles and job types.
& ``Certain jobs could disappear faster than new roles are created, which might cause short-term instability in some sectors. Without strong retraining programs, that gap could leave parts of the workforce struggling to keep up.''
& 6.10\% \\ \cline{1-4}

AI in Creativity
& Participants discuss AI’s impact on creative professions, including discussions about quality and job security.
& ``I like ai art and ai video, creating a lot getting followers, make money. plus AI give me better advice than humans.''
& 5.20\% \\ \cline{1-4}

Economic Inequality and AI
& Concerns are raised about AI exacerbating economic inequality and limiting job access for certain groups.
& ``The wealth gap will continue to expand and there will be a lot of people left behind and will struggle financially.''
& 4.60\% \\ \cline{1-4}

Mixed Feelings About AI
& Participants express both positive and negative feelings about AI's impact on their careers and the economy.
& ``I think AI could have both positive and negative effects. On the positive side, it might reduce paperwork and help with diagnostics, making my job more efficient and less stressful. But there’s also a chance that certain administrative or routine roles could be reduced, which could affect job security. Overall, I’m cautiously optimistic as long as we adapt and upskill.''
& 3.70\% \\

\hline
\multicolumn{4}{l}{\textbf{ Personal Strategies to Cope with AI's Economic Impact (5,670)}} \\ \hline

Rethink Career Path
& Participants worry about job displacement in an AI-driven world and think about career changes.
& ``They will probably need to find another profession, just like people in the logging industry had to deal with in the 1980s.''
& 20.10\% \\ \cline{1-4}

New Skill Development and Learning
& Emphasis on acquiring new skills and ongoing education to adapt to AI advancements.
& ``I have watched a few YouTube videos about AI and the way it is planned on being used in the future.''
& 16.90\% \\ \cline{1-4}

Coping with Uncertainty
& Strategies and feelings of uncertainty regarding the future impact of AI on lives and careers.
& ``I'm sure others will. I'm not sure if I will. It all depends on how much preparation is required, how expensive the skills are to build. If I absolutely need them to make a comfortable wage and I can afford to build them, it's likely I'd go for it.''
& 13.60\% \\ \cline{1-4}

Adaptation Strategies
& Strategies individuals are implementing to adapt to AI in various contexts.
& ``At work, I try to stay aware of any new technologies being introduced and think about how I can work alongside them rather than be replaced by them. In my personal life, I’ve started using AI tools to help with tasks like organizing information, learning new skills, and even exploring creative projects.''
& 12.20\% \\ \cline{1-4}

Utilizing AI for Efficiency
& Discussion on how AI tools can enhance productivity and streamline tasks.
& ``I used AI for generating market summaries which cuts lots of time for me (18787).''
& 11.30\% \\ \cline{1-4}

Resistance to AI Integration
& Skepticism or resistance toward adoption and integration of AI technologies.
& ``I am very against using generative AI. AI should not be used to think for people. Also, I am very concerned by the vast amount of resources AI uses.''
& 8.80\% \\ \cline{1-4}

Balancing AI and Human Skills
& Need to maintain human skills and critical thinking alongside AI integration.
& ``I see AI as a tool that I can add to my tool box. It forces me to be MORE creative and MORE innovative.''
& 2.20\% \\

\hline
\multicolumn{4}{l}{\textbf{ People's Perceptions of Government’s Role in AI-Driven Economic Change (4,402)}} \\ \hline

Need for Government Regulation
& Participants emphasize the necessity of government regulation for social changes.
& ``I think government needs to regulate AI to limit as many job losses as possible.''
& 36.20\% \\ \cline{1-4}

Ethical Use/Development of AI
& Emphasis on ethical considerations and guidelines for AI usage and development.
& ``Educating people on how to properly use AI, how to set limits, and some of the consequences of failing to do so would be great.''
& 15.10\% \\ \cline{1-4}

Education and Training Initiatives
& Calls for government-funded education and training programs to prepare the workforce.
& ``Job transition support such as funding large-scale retraining programs and apprenticeships.''
& 14.90\% \\ \cline{1-4}

Job Transition Support
& Need for targeted workforce retraining and income support programs.
& ``The concrete step the government could take is investing in targeted workforce retraining and income support programs.''
& 11.50\% \\ \cline{1-4}

Universal Basic Income (UBI)
& Advocacy for UBI as a response to job displacement from AI.
& ``If it doesn't set up something akin to Universal Basic Income, I assume the economy will be exponentially worse.''
& 8.50\% \\ \cline{1-4}

Skepticism Towards Government Involvement
& Skepticism about the government’s ability to manage AI-related issues effectively.
& ``No political action is needed. I think people and companies will work things out.''
& 8.10\% \\
\hline
\caption{Study 2 Generated Report: Interview Data}
\end{longtable}

\end{document}